\documentclass[preprint]{aastex}        	

\usepackage{natbib}
\usepackage{apjfonts}	

\slugcomment{VERSION 14 January 2003; Accepted to the Astronomical J.}
\shorttitle{Wide-Field NIR Imaging of IC~348}
\shortauthors{Muench et al.}

\newcommand{\solarmass}{\ensuremath{ \mathnormal{M}_{\Sun} }}
\newcommand{\jupmass}{\ensuremath{ \mathnormal{M}_{Jup} }}
\newcommand{\mass}{\ensuremath{ \mathnormal{M} }}
\newcommand{\chisq}{ \ensuremath{ \chi^{2} }}
\newcommand{\av}{\ensuremath{ \mbox{A}_{V} }}
\newcommand{\ak}{\ensuremath{ \mbox{A}_{K} }}
\newcommand{\Ks}{\ensuremath{ {K}_{s} }}
\newcommand{\mj}{\ensuremath{ \mbox{m}_{J} }}
\newcommand{\mh}{\ensuremath{ \mbox{m}_{H} }}
\newcommand{\mk}{\ensuremath{ \mbox{m}_{K} }}

\newcommand{\Mv}{\ensuremath{ \mbox{M}_{V} }}
\newcommand{\vvi}{\ensuremath{ V\,/\,V-I }}

\newcommand{\halpha}{\ensuremath{ \mbox{H}\alpha }}

\begin{document}

\title{
A Study of the Luminosity and Mass Functions of the Young IC~348 Cluster \\
using FLAMINGOS Wide-Field Near-Infrared Images
}

\author{
A. A. Muench\altaffilmark{1,4,5},
E. A. Lada\altaffilmark{1,4}
C. J. Lada\altaffilmark{2,4},
R. J. Elston \altaffilmark{1,4},
J. F. Alves\altaffilmark{3,4}, \\
M. Horrobin\altaffilmark{1,4},
T. H. Huard\altaffilmark{2,4},
J. L. Levine\altaffilmark{1,4},
S. N. Raines\altaffilmark{1,4}, and
C. Roman-Zuniga\altaffilmark{1,4},
}
\altaffiltext{1}{Department of Astronomy, University of Florida, 
     Gainesville, FL 32611}
\altaffiltext{2}{Harvard-Smithsonian Center for Astrophysics, 
     Cambridge, MA 02138}
\altaffiltext{3}{European Southern Observatory, Karl-Schwartzschild-Strasse 2, 
     8574 Garching Germany}
\altaffiltext{4}{Visiting Astronomer, Kitt Peak National Observatory.
     KPNO is operated by AURA, Inc.\ under contract to the National Science
     Foundation.}
\altaffiltext{5}{present address: SIRTF Science Center, California Institute 
     of Technology, Pasadena, CA 91125 USA}

\begin{abstract}

We present wide-field near-infrared $(JHK)$ images of the young,
$\tau=2\,\mbox{Myr}$ IC~348 cluster taken with FLAMINGOS.
We use these new data to construct an infrared census of sources,
which is sensitive enough to detect a $10\,\jupmass$~brown dwarf seen
through an extinction of $\av\sim7$.
We examine the cluster's structure and relationship to the molecular
cloud and construct the cluster's $K$ band luminosity function.
Using our model luminosity function algorithm we derive the cluster's 
initial mass function throughout the stellar and substellar regimes 
and find that the IC~348 IMF is very similar to that found for 
the Trapezium Cluster with both cluster IMFs having a mode 
between $0.2 - 0.08\,\solarmass$.
In particular we find that, similar to our results for the Trapezium,
brown dwarfs constitute only 1 in 4 of the sources in the IC~348 cluster.
We show that a modest secondary peak forms in the substellar IC~348 KLF, 
corresponding to the same mass range responsible for a similar KLF peak
found in the Trapezium.
We interpret this KLF peak as either evidence for a corresponding 
secondary IMF peak at the deuterium burning limit, or as arising from 
a feature in the substellar mass-luminosity relation that is not predicted
by current theoretical models. 
Lastly, we find that IC~348 displays radial variations of its sub-solar
$(0.5 - 0.08 \solarmass)$ IMF on a parsec scale.
Whatever mechanism that is breaking the universality of the IMF on
small spatial scales in IC~348 does not appear to be acting upon
the brown dwarf population, whose relative size does not vary with
distance from the cluster center.

\end{abstract}



\section{Introduction}
\label{sec:intro}

For the past few decades, advances in near-infrared (NIR) detector technology
have provided new and exciting discoveries of young clusters and star
forming regions.  For example, the partially embedded cluster IC~348,
which is the focus of this paper, was first observed in the infrared using
a single channel photometer \citep{strom74}. However, such early observations
were severely limited in both sensitivity and angular resolution, allowing for
the detection of only the brightest embedded cluster members. For example,
\citeauthor{strom74} observed 25 possible stellar members of the IC~348
cluster with $K\le$10.5 while discovering a few bright but heavily embedded
IR sources.

The development of NIR infrared imaging arrays $\sim$ 15 years ago,
greatly enhanced our ability to study young clusters by not only improving
the sensitivity and angular resolution of NIR studies but also the areal
coverage.  For the first time, large scale ($\sim$ square degree) surveys
of molecular clouds \citep[e.g., Orion B GMC; ][]{lada91} and in particular
of young clusters such as IC~348 \citep{ll95} were conducted. Such
surveys revealed the presence of hundreds of young stars over a wide mass
range contained within the young clusters, increasing the number of
stellar sources in IC348 by a factor of 15 and increasing the depth by a
factor of 40 over earlier studies.  While these early NIR imaging surveys
were clearly a vast improvement over surveys that had used single channel
photometers, they were not without their limitations. In particular they
required obtaining large mosaics containing tens to thousands of
individual frames per wavelength band to cover the desired star forming
area.  For example \citeauthor{ll95} obtained twenty four separate
fields to cover the central parsec of IC~348.  Consequently there was
often a trade off between survey depth and area covered and, therefore
 in the case of many young clusters, obtaining a complete census of
its members was still unattainable.  

The most recent advancements in NIR detector technology have permitted the
development of sensitive, wide-field infrared cameras such as FLAMINGOS
\citep[][Elston et al. 2003, in preparation]{elstonflam98}, allowing more
than an order of magnitude improvement in areal coverage and depth. 
In terms of young cluster studies, these improvements now enable
us to obtain a complete census of stars of all masses,
even deeply embedded ($\av\,\sim\,5-10$) young brown dwarfs
with masses as small as 10 Jupiter masses, over the complete spatial
extent of the cluster. Observing the full angular extent of young
clusters may be critical for obtaining a complete stellar census since
recent theoretical studies show that mass segregation (primordal or dynamical)
may be present in even the youngest embedded clusters \citep{bonn01,kah00}.
Thus, for the first time we are poised to study the complete populations of
young star clusters to below the hydrogen burning limit rather than 
focusing only on the stellar population of only the clusters' cores.

Since the \citeauthor{ll95} NIR survey, IC~348 has been the target of
a number of wide-field optical \citep{luh99b}, $\halpha$~\citep{herbig98}
and X-ray imaging \citep{pre96} studies, and all have reinforced the ideas
that  IC~348 is partially embedded at the edge of the Perseus GMC and 
IC~348 is spatially extended on the sky beyond its central core \citep{ll95}.
With our current wide-field NIR image we are able to survey 
an area encompassing nearly all the boundaries of these past surveys 
while simultaneously extending the  \citeauthor{ll95} NIR survey depth 
by nearly 3 magnitudes. Thus, we are able to detect very low
mass sources ($0.01 \solarmass$) over a much larger cluster volume
(larger extinction) than any prior survey.  We use the results of our
infrared census to examine the cluster's structure, reddening and
relationship to the Perseus Molecular Cloud (Section \ref{sec:images}) 
and construct the cluster's differential $K$ band luminosity function
(Section \ref{sec:lf}).  In Section \ref{sec:imf} we combine our deep 
survey, a spectroscopically determined age estimate for IC~348 from
\citet{herbig98} data, and our improved luminosity function modeling
techniques \citep[][hereafter, Paper I]{aam02} to derive the IMF of
IC~348 across the stellar and sub-stellar regimes.  \citet{ll95} found
that the turnover in the IC~348 KLF can be modeled by a similar peak
and turnover in the underlying IMF; we expand on their finding by showing
that the composite cluster IMF forms a mode near 0.1 \solarmass, turns over 
and declines throughout the brown dwarf regime down to near $30\,\jupmass$.
Our wide-field imaging also allows us to derive the cluster's stellar and
sub-stellar IMF over a much larger area than similarly sensitive surveys,
which have concentrated on the cluster's central
core \citep[][hereafter, NTC00]{luh98, ntc00}. 
From this analysis, we compare our results for IC~348 and the
Trapezium (Paper I) and discuss in Section \ref{sec:discuss} the
impact of spatial or statistical IMF variations on meaningful 
comparisons of the IMFs for different clusters.


\section{Wide-Field Near-Infrared Images of IC~348}
\label{sec:images}

\subsection{FLAMINGOS Observations}
\label{sec:images:obs}

We obtained wide-field near-infrared images of the IC~348 cluster using the 
FLoridA Multi-object Imaging Near-IR Grism Observational Spectrometer
\citep[FLAMINGOS, ][]{elstonflam98}
on the 2.1m telescope at the Kitt Peak National Observatory, 
Arizona (USA) during December 2001 and February 2002. The FLAMINGOS instrument 
employs a 2K HgCdTe ``HAWAII-2'' imaging array, which when configured on the 
2.1m Kitt Peak telescope yields a $20.5\,\arcmin\,\times\,20.5\,\arcmin$ field
of view with a derived plate scale at $K$ band of $\sim0.608\,\arcsec/\mbox{pixel}$.

\placetable{tab:obs}

On both 14 December 2001 and 08 February 2002 dithered sets of IC~348 images 
were obtained with FLAMINGOS in the $J$ $(1.25\;\micron)$, the $H$ $(1.65\;\micron)$
and the $K$ $(2.2\;\micron)$ (or~$\Ks\,--\,\,2.162\;\micron$) bandpass.
These image sets were obtained within a narrow window of time (1.5 - 2.5 hrs) 
and a restricted range of airmass ($\sec\mbox{z} < 1.4$).  We list the details 
of these observations in Table \ref{tab:obs}.  Briefly, we employed a large 
number (15-50) of short (20-60 sec) non-repeating dithers to yield total 
integration times of approximately 14-24 minutes, depending upon the bandpass.  
Conditions on both nights appeared photometric, with no apparent cloud cover, 
stable background sky counts, and seeing estimates between $1.6 - 1.9\arcsec$
on 14 December 2001 and $1.5 - 1.7\arcsec$ on 08 February 2002.

Two wide-field off-cluster images were similarly imaged in the $K$ or $\Ks$~bands
on 17 December 2001 and 11 February 2002. These two non-overlapping regions lie 
approximately 1 degree east of IC~348, along a line of constant galactic 
latitude with the young cluster.  Their equatorial field centers were:
1) $03^{h}48^{m}21\fs9$; $+31\degr38\,\arcmin06\farcs7$ (J2000); and
2) $03^{h}48^{m}19\fs4$; $+31\degr08\,\arcmin34\farcs7$ (J2000).
Additional details of these off-field observations are also listed in
Table \ref{tab:obs}.

We reduced the sets of dithered frames with the April 2002 version of the
FLAMINGOS data reduction pipeline (Elston et al 2003, in preparation).
Local sky flat-fields were used for $J$ and $H$ bands,
while dome flats were used for $K$ or $\Ks$~bandpass.  
Briefly, the FLAMINGOS data reduction pipeline is based upon a two-pass routine 
with object masking during the second pass that permits the creation of star-free 
median sky frames from the target images while following standard techniques for the
reduction of near-infrared data. To take advantage of the relatively large number of
dithers in our datasets, the pipeline employs the drizzle IRAF routine \citep{driz02} 
to allow for linear sub-pixel image reconstruction during the final combination 
of the dithered frames. In summary, 6 reduced cluster images (2 at each 
bandpass) were obtained, in addition to two $K$ band off-field images.

\placefigure{fig:jhk_image}

In Figure \ref{fig:jhk_image} we display a (false) color composite,
near-infrared image of IC~348 using the $J$, $H$, and $\Ks$~FLAMINGOS images from 
our February 2002 observations.  A number of interesting cluster features are 
outlined by low-level nebulosity.  These include the cluster core, which 
displays deep red nebulosity suggestive of somewhat higher extinction, and 
the interface between the IC~348 cluster and the molecular cloud all along 
the southern edge of the image. This interface region includes numerous 
signposts of very recent star formation including the HH-211 infrared jet
 \citep{mrz94}, the enshrouded NIR source deemed the ``Flying Ghost'' nebula
 \citep{strom74, flyghost}, a dark lane suggestive of a flared, edge-on
disk-like structure, and a number of bright infrared sources detected 
only in the $K$ band.

A few image artifacts can also be seen in this figure. These include geometric 
distortions of the stars in the northeast corner of the image, red and green 
glint features due to internal reflections along the southern edge of the cluster, 
and coma-like ghosts south-southwest of the cluster center, resulting from a
long time constant in one amplifier of the HAWAII-2 array.  On the other hand, 
the alternating blue-green-red source located south-southwest of the cluster center 
is not an image artifact, but instead is the asteroid 545 Messalina.

\subsection{The Infrared Census}
\label{sec:images:census}

\subsubsection{Photometry and Calibration}
\label{sec:images:phot}

We characterized each reduced image by deriving estimates of the FWHM of the
stellar point spread function and the pixel-to-pixel noise in the background 
sky using the IMEXAMINE IRAF routine, although the pixel-to-pixel noise is 
correlated in drizzled images. The resulting seeing estimates and $5\sigma$ 
detection limits are listed in Table \ref{tab:obs}. We find that the slightly 
better seeing and longer exposure times of the February data yielded detection 
limits approximately 1 magnitude fainter than the December observations.
Sources were initially identified on each reduced image using the stand-alone 
S-Extractor package \citep{bertin96}. In an iterative fashion, accurate 
centroids were calculated for the detected sources using the CENTER IRAF 
routine, and marked on the reduced images, after which the images were 
manually searched to identify false detections or to add sources missed 
near bright stars. The source lists for each of the 6 on-cluster images
were then cross-correlated to identify and check those sources not appearing
on all the images.

Multi-aperture photometry was performed on the sources using the APPHOT IRAF 
package and the instrumental magnitudes were corrected out to the beginning 
of the sky annulus using photometric curves of growth calculated from
$\sim20-30$ bright stars using the MKAPFILE IRAF routine.  From the corrected 
multi-aperture photometry, we chose to use the smallest beamsize that 
simultaneously gave the most consistent photometry when compared to larger
apertures.  Because of a spatially varying PSF due to geometric distortions 
present in the final images, we resorted to using a rather large aperture
(radius = 5 pixels; beamsize = $6\arcsec$), yielding aperture corrections 
typically of order $\lesssim\,0.06$ magnitudes. Absolute calibration of the 
instrumental photometry was performed using zeropoint and airmass corrections 
derived from \citet{persson98} standard stars, which were observed on the same 
night as the targets. For secondary calibration, we calculated and removed 
median zeropoint offsets of order $\sim0.04-0.08$ magnitudes between our absolute 
photometry and the 2MASS photometric system. Because the natural system of 
our photometry is not known, the observed colors of our sources have not been
transformed to the 2MASS system. 
Complete source lists for the FLAMINGOS IC~348 cluster observations presented
in this paper, including equatorial positions and aperture photometry from the
December 2001 and February 2002 datasets, will be made available electronically
at the FLAMINGOS website at the University of Florida 
      \footnote{\url{http://www.flamingos.astro.ufl.edu/}}.
Final products of the associated NOAO Survey project, ``Toward a Complete
Near-Infrared Spectroscopic and Imaging Survey of Giant Molecular Clouds,''
will be made available through the NOAO Science Archive
      \footnote{\url{http://archive.noao.edu/nsa/}}.

\subsubsection{Accuracy}
\label{sec:images:err}

We quantified our photometric accuracy by directly comparing the photometry from
the December and February images and by matching our data to the 2MASS catalog.
We list the results of our photometric comparison to 2MASS in Table \ref{tab:2mass}.
This table includes the magnitude range of the comparisons based on the 
FLAMINGOS saturation and 2MASS photometric limits, the secondary calibration
offsets applied in Section \ref{sec:images:phot}, and the $1\sigma$ dispersion 
between the datasets.

\placetable{tab:2mass}

While the 17 December off-field data display a dispersion of $\sim3\%$ at
K band relative to the 2MASS catalog, we found $JHK$ dispersions in the
14 December on-cluster data that were larger than expected from purely photometric
errors with the $K$ band scatter $4\%$ larger in the on-cluster frame than that
found in the off-cluster frame taken three days later under similar sky
conditions.  We attribute some of this additional scatter to the intrinsic 
variability of young PMS stars in IC~348 \citep{herbst00}, which in the 
infrared typically peaks at $J$ band \citep{chs01} (i.e., for our
IC~348 data $\sigma_{K}=0.05$ versus $\sigma_{J}\sim0.07$ mag).

Between 03 - 11 February 2002, oil from the primary mirror cell contaminated
the FLAMINGOS dewar window. This contamination had a negligible effect on our $J$
and $H$ band photometry since these data were flat-fielded with local, contemporaneous
sky flats.  However, the $K$ data were flat-fielded using non-contemporaneous dome
flat-fields and display a $4-6\%$ increase in the photometric scatter relative to the
FLAMINGOS data obtained in December and to the 2MASS catalog. This scatter does not
affect the analysis or conclusions of this paper.

\placefigure{fig:cmag_all}

\subsubsection{Results}
\label{sec:images:results}

After considering the different saturation, detection and noise limits of the
December and February IC~348 FLAMINGOS data, we produced a single working source list
with photometry drawn from both FLAMINGOS datasets.  We chose to use the
December $K$ band data down to $\mk\sim15.5$.  Below this value we transitioned to the
intrinsically deeper February \Ks~data.  In the $J$ and $H$ bands we averaged the two
datasets between the saturation and the detection limits of these respective catalogs,
recording the $1\sigma$~standard deviation between the two observations.
For the brightest 39 stars, which were saturated in one or more bands on all of
our images, we substituted the 2MASS photometry for our FLAMINGOS photometry.
Finally, for the faintest objects ($\mj > 18.5$; $\mh > 17.5$; $\mk > 17.0$) we
used smaller aperture photometry (radius = 3.5 pixels, aperture correction =
$\sim -0.12$ mag).

In Figure \ref{fig:cmag_all} we present the infrared color-magnitude diagrams 
($J-H$ vs $H$ and $H-K$ vs $K$) for the FLAMINGOS IC~348 region without filtering for 
photometric error. We compare the distribution of sources in these diagrams to
the location of pre-main sequence isochrones taken from the \citet{dm97} and
\citet{bcah98} evolutionary models.  For these comparisons we assume a cluster 
mean age of 2 Myr and a distance of 320 pc (see Sections \ref{sec:imf:sfh} 
and \ref{sec:imf:dist} for further discussion of these parameters). 
For these cluster parameters we find that we are sensitive to a 2 Myr,
$80\jupmass$~object at the hydrogen burning limit seen through
$\sim30$ magnitudes of extinction or a $10\jupmass$~brown dwarf near
the deuterium burning limit seen at an
$\av=7$~\citep[using the][theoretical models]{bur97}. 
Three general characteristics of the sources projected towards our IC~348
FLAMINGOS region are clearly seen: 
1) The color-magnitude diagrams indicate a cluster region having only modest
   reddenings relative to other star forming regions, with the vast majority of
   the sources having $\av < 7$; 
2) There is a density of sources between $10 < \mk = \mh < 15$ that is closely
   outlined by the PMS isochrones and represents a range of magnitude space
   that appears dominated by likely cluster members;
3) Below $\mk = \mh = 15$ the color-magnitude distribution appears to become
   dominated by field-stars (and likely galaxies) as indicated by a rapid
   increase and steady broadening in the density of sources on these plots.
All of these features were seen in individual datasets (December vs February)
and were not modified if we changed any of the parameters for merging the data
(e.g., the use of $K$ instead of $\Ks$ data for the faintest objects).

\placefigure{fig:jhk}

In Figure \ref{fig:jhk}, we display the infrared $H-K$ vs $J-H$ color-color diagram 
for sources in the IC~348 region.  We include sources most likely to be cluster 
members by using characteristics (2) and (3) of the color-magnitude diagrams listed
above to apply a somewhat arbitrary $\mk=15$ magnitude limit.
The ``cluster sample'' resulting from this magnitude cut should be complete for
unreddened 2 Myr brown dwarfs down to $0.025\solarmass\,(\sim25\jupmass)$  
and for $0.04\solarmass\,(\sim40\jupmass)$ brown dwarfs seen through $\av\,\sim\,7$.

There are 580 sources with $\mk<15$ of which 563 have $JHK$ photometry; we display all
of these in the color-color diagram without filtering for error. Again, it is clear
that the cluster region is only marginally reddened with the nearly all of the
sources having $\av < 7$.  The overwhelming majority of the sources fall within a
locus bounded by the reddening vectors for the giant branch and the
tip of the M dwarf sequence. While one source has NIR colors significantly to
the left of the reddening band, $66$ sources have infrared colors which lie to
the right of the reddening vector for M9 dwarfs \citep[$H-\Ks\,=\,0.46$;][]{kirk00}.
These sources fall into a region of infrared excess in the color-color diagram,
which because IC~348 is a very young cluster are considered to be likely cluster
members with optically thick, circumstellar disks. 
Finally, 42 of these sources exhibit infrared excesses
greater than their $1\sigma$ photometric noise, and we note that these filtered
sources span the entire range of H band magnitude both above and below
the expected luminosity of the hydrogen burning limit.
With these basic observational results we examined the structure of the IC~348 cluster.

\subsection{Cluster Structure}
\label{sec:images:struc}

For our subsequent analysis of the luminosity and mass function of IC~348 we 
wished to select that portion of the FLAMINGOS wide-field image that provides the
best sampling of the overall cluster. Studying a region somewhat similar in size
to the FLAMINGOS area, \citet{ll95} used a surface density analysis to show that
IC~348 could be broken into 9 apparent sub-clusterings spread across their
survey region.  Although subsequent wide-field \halpha~\citep{herbig98}
and optical \citep{luh99b} surveys covered areas similar to LL95, 
most studies of the IMF of IC~348 have concentrated on the central
LL95 sub-cluster, IC348a \citep[][; NTC00]{herbig98,luh98} and have not included
the other LL95 sub-clusters.  While existing optical and \halpha~studies
may systematically underestimate the cluster size due to extinction or miss
members that do not display \halpha~emission, they have confirmed that the 
cluster is in fact spread over an area larger than the IC348a region.  Using the
deeper wide-field near-infrared imaging provided by our FLAMINGOS observations,
we re-investigated the structure of the IC~348 cluster by calculating the
cluster's radial profile and by examining the spatial distribution of sources.

\subsubsection{Cluster Radius}
\label{sec:images:radprof}

In Figure \ref{fig:rad_prof} we construct the radial profile of sources in
the FLAMINGOS IC~348 region, centering on the IC~348a sub-cluster.
We use only those sources for which $\mk<15$ and calculate the surface density in stars
per square degree using both annuli of equal width and annuli having constant areas. 
We compare the resulting radial profiles to the field star surface density
calculated from our off-field data. Clearly the cluster exceeds the
unreddened background surface density over most of the FLAMINGOS region.
Specifically, the cluster extends over an area considerably larger than the
IC348a sub-cluster, whose radius was given as $0.47$ pc ($5.05\,\arcmin$)
in LL95 and over an area larger than the radius of $4\,\arcmin$~calculated
by \citeauthor{herbig98}. The cluster appears to dip to the unreddened
background surface density at a radius of $\sim10-11\,\arcmin$; however,
since this radius is also where the profile begins to clip the edge
of the survey region, we cannot confidently rule out a larger cluster
radius using the current FLAMINGOS image.  At the distance of IC~348,
this translates to a cluster radius of $\sim1\,\mbox{pc}$, similar to
the effective radius of $1.19\,\mbox{pc}$ derived by \citet{carp00}.

\placefigure{fig:rad_prof}

To provide a means for comparing our study to that of other authors,
we separated our FLAMINGOS cluster region into two sub-divisions based upon
the radial profile and assigned these two sub-regions primarily
functional names, e.g., the cluster ``core'' sub-region with a
radius = $5\,\arcmin$ and the cluster ``halo'' region between
the cluster ``core'' and a radius of $10.33\,\arcmin$, corresponding to the
largest unclipped radius permitted by the current FLAMINGOS survey.
The core region approximately represents the IC~348a sub-cluster but is slightly
larger than that area studied by \citet{luh98} and NTC00, while the
halo region covers an area approximately 3.3 times that of the core and 
encompasses the stars contained in the sub-clusters designated {\em b-i} by LL95.
Naturally, we are viewing a three-dimensional cluster in projection, such
that true ``halo'' sources will also be present in the cluster's ``core''
sub-region, and therefore will somewhat blur the meaning of our division of the
cluster by area.

Although our construction of the cluster's radial profile allowed us to
divide the spatial extent of IC~348 within our survey region, we found
that it follows neither a simple $\mbox{r}^{-1}$ nor a King profile.
We fit both analytical profiles to the cluster's constant area annuli profile,
varying the radial extent of the fits, and for the King profile, allowing
both the core and tidal radii to vary freely. Neither analytic profile
provided reasonable $\chisq$~fits to the entire cluster profile.
As we display in Figure \ref{fig:rad_prof}, the tail of the cluster profile,
(i.e., the halo sub-region), is much flatter than either of the analytic
profiles, and this is likely the reason \citeauthor{herbig98} derived a smaller
cluster radius.  One interpretation of these fits is that the LL95 sub-clusters,
which constitute the cluster's halo, are actually separate entities,
rather than being statistical fluctuations on a $\mbox{r}^{-1}$ cluster profile. 

Our construction of the IC~348 radial profile and the subsequent analytical
fits likely suffer from two problems.
First, we use circular annuli, while, as we will show in Figure
\ref{fig:spatial}b, the cluster is elliptically elongated in the N-S direction.
Second, we rely upon an empirical estimate of the field-star surface density
from a region near to IC~348 but that, in principle, could fluctuate between
this location and the background relevant to IC~348. Indeed, the surface
densities of the two off-cluster locations fluctuate more than expected purely
from counting statistics despite their similar galactic latitudes. 
This spatial fluctuation, however, is exceedingly small relative to
the excess surface density of the cluster halo and will not likely affect
the cluster boundary we derive.
On the other hand, the obscuration of background field stars by the parental
molecular cloud will lower the expected field star surface density and will
affect the derived cluster radius, as we illustrate in
Figure \ref{fig:rad_prof}.  Reddening the background by the average
extinction in the IC~348 region($\av=4$; see Section \ref{sec:images:redden})
only expands the cluster radius by $1-2\,\arcmin$, however, and the cluster
probably does not exceed a radius of $15\,\arcmin$ (1.4pc), which is a boundary
traced very clearly by wide-field X-ray detections \citep{pre96}.

\placefigure{fig:spatial}

\subsubsection{Spatial Distribution of Sources}
\label{sec:images:spatial}

With the purpose of characterizing our two cluster sub-regions, we examined
the spatial distribution of the IC~348 FLAMINGOS sources, plotting them using
separate symbols for different luminosity ranges in Figure \ref{fig:spatial}a.

By segregating the two sub-regions, we find, for example, that there are as many
relatively bright sources $\mk<10$ (i.e., with $\gtrsim\,1\,\solarmass$) in the 
cluster core as in the cluster halo. Since there are only a handful of early (B or A) 
type stars in IC~348, there is no statistically meaningful way to show that the
3 BA stars in the cluster core are ``segregated'' relative to the 2 A stars and
the B0III star $\omicron$ Persei in the cluster halo. 
On the other hand, the faintest sources ($\mk>15$) do show some spatial
variations, with a sharp decrease in these sources along the cluster's
southern edge, outlining the interface of the cluster with the
molecular cloud.

In Figure \ref{fig:spatial}b, we display a surface density plot of IC~348
from the 580 sources with $\mk<15$  used to construct the cluster's radial
profile (see Figure \ref{fig:rad_prof}) and smoothed with a box filter
at the Nyquist sampling. The box filter had a width of $200\arcsec$ or
an area roughly equivalent to the area of the annuli used in the radial profile.  
At this resolution, only a smooth N-S elongated cluster is seen with
no significant sub-clusterings.  We confirmed, however, that
if we use the same spatial resolution as the
\citeauthor{ll95} study ($90\,\arcsec$), we recovered most of the sub-clusters
identified by LL95 and that now lie in the cluster's halo.
Further, these surface density contour maps display no apparent correlation
to the location of the cluster-cloud interface region at either resolution.

Lastly, we examined the locations of the 66 sources displaying NIR excess. 
As was found by LL95, the majority of these sources lie outside of
the cluster core.  Moreover, the surface density of these NIR excess sources
appears to increase toward the southern interface with the Perseus Molecular
Cloud, opposite to the trend in the distribution of the faintest objects.
These patterns suggest that many of the NIR excess sources are correlated to
and likely embedded within the molecular cloud, and may be associated with
the most recent star formation events in the IC~348 region.

\subsection{Cluster Reddening Properties}
\label{sec:images:redden}
 
We also used the sources observed within the IC~348 region as line of sight 
probes of the parental molecular cloud, allowing us to characterize the 
reddening that would be seen towards cluster members or background field-stars.
Building upon methods developed in \citet{lada94} and \citet{alv98} and our recipe(s)
described in \citet{aam02}, we calculated individual extinction estimates for
each source by de-reddening the sources' infrared colors to a locus of assumed
intrinsic colors in the $(H-K)$ vs $(J-H)$ color-color diagram.
These extinction estimates were used to create reddening maps of the
cluster region and show that IC~348 is primarily foreground of the
remnant molecular material. 
Further, we binned these individual \av\ estimates into extinction probability
distribution functions (hereafter known as EPDFs), which we will use when estimating
the number of interloping field-stars and when calculating luminosity function models
to interpret the observations.

\subsubsection{Extinction Estimates and Reddening Maps}
\label{sec:images:avmaps}

We calculated individual $\av$~ for sources falling into two different luminosity
ranges, dividing them into sets that are likely dominated by either cluster members
or background field-stars.  We separated the ``bright -- cluster'' and
``faint -- background'' samples based upon their location in the $(H-K)$
vs $K$ color-magnitude diagram relative to the reddening vector of a source
at $\mk = 15$ ($H-K = 0.35$).  The ``faint'' sample was also limited to
objects brighter than a reddening vector for a source at $\mk = 17, (H-K = 0.4)$.

\placefigure{fig:av_map}

The two magnitude samples were de-reddened back to different loci of assumed
intrinsic colors.  The ``bright'' sources were assumed to be young PMS stars
with masses as small as $\sim\,0.03\,\solarmass$. $95\%$ of these have $JHK$
colors and were de-reddened back to the classical T-Tauri Star (cTTS) locus
in the $JHK$ color-color diagram \citep[][ slope = 0.58;
$J-H$ intercept = 0.52]{mch97}, while the remaining $5\%$ were assigned an
intrinsic $H-K$ color = 0.5 and $\av$~estimates derived.  The ``faint'' sources
were assumed to be dominated by field M dwarfs and those with $JHK$ colors
($\sim85\%$) were de-reddened back to a linear approximation of the M dwarf
branch in the color-color diagram (from K6 to M9, slope = 0.16; $J-H$ intercept =
0.61)\footnote{These cTTS and M dwarf approximations result in nearly all field
giants being de-reddened to the colors of the K1 - K2III spectral class.},
while the remaining $15\%$ lacking $J$ band were assigned an intrinsic $H-K = 0.16$.

In Figure \ref{fig:av_map} we present the resulting extinction maps derived
from these two samples.  Both maps clearly define the location of the
cluster-cloud interface along the region's southern border, while they also
outline a NE-SW band of reddening that passes through the cluster's core,
similar to the cluster's N-S elongation.  One straightforward conclusion
from the spatial variations in either reddening map
is that the IC~348 region cannot be characterized by a single mean
$\av$~value.  Although the two $\av$~maps are physically very similar, the
reddening seen by the background stars is in general larger than that
foreground to the cluster members, indicating that the remnant molecular
cloud lies primarily behind the cluster core rather than in front of it.

\subsubsection{Extinction Probability Distribution Functions (EPDFs)}
\label{sec:images:epdfs}

In Figure \ref{fig:hk_av} we plot the normalized histograms of the ($H-K$) color
and of $\av$~for sources in the two magnitude ranges defined in the previous
section and separate them further by cluster sub-region. As expected from the
$\av$~maps, the faint sources' color and $\av$~distributions are broader and
redder than the bright, likely cluster stars, although the EPDFs of the likely
cluster members in both regions appear somewhat similar. 
Applying a two-sided Kolmogorov-Smirnov test to the $\av$ values derived
from bright and faint sources in the cluster core, we found that it is unlikely
that they are drawn from the same $\av$~distribution, having a KS probability
of only 0.00024.  This is in contrast to the halo sub-region, where the bright
and faint stars have a 0.23 probability of being drawn from the same $\av$~
distribution.  Similarly the bright stars in the core and halo have a
0.043 probability of being drawn from the same distribution while the
faint stars in both regions cannot, ruled out at the $2.0000 \times 10^{-8}$
confidence. Taken together, the EPDFs and these statistical tests
support two basic conclusions about the reddening seen towards IC~348:
1) there is a measurable difference in the reddening seen
by the background stars between the core and halo sub-regions, owing
to the cluster lying in front of substantial molecular material
(also see Figure \ref{fig:av_map}b);
2) the bright, likely cluster members of the core and halo appear to
have fairly similar reddenings, despite the projection of the entire
region onto various pieces of the Perseus GMC. 

\placefigure{fig:hk_av}

The normalized $\av$~histograms (i.e., EPDFs) are generally
skewed to $\av<5$ and are quite non-gaussian, again indicating that a 
single $<\av>$ value is inappropriate to describe the cluster reddening. 
On average we find reddenings to the background stars of
$<\av>_{core}\;\sim\;7.2\;(1\sigma = 6.3,$ avg. deviation = 4.6, median = 5.3)
and
$<\av>_{halo}\;\sim\;4.6\;(1\sigma = 4.3,$ avg. deviation = 3.0, median = 3.4),
while to the likely cluster members these averages are
$<\av>_{core}\;\sim\;4.9\;(1\sigma = 4.2,$ avg. deviation = 2.9, median = 3.8)
and
$<\av>_{halo}\;\sim\;4.2\;(1\sigma = 4.0,$ avg. deviation = 2.6, median = 3.3).
These latter averages are roughly the same as the $\ak\;\sim\;0.5$ assumed
by LL95 for the entire IC~348 region. More meaningful comparisons can be made
to those studies that have used spectroscopic observations to derive individual
$\av$ estimates. Our $<\av>$ estimates are consist with the $<\av>\;=\;2.8$
derived by \citeauthor{herbig98} over a large cluster area.
In the core, we compared our $\av$ estimates to the $\ak$ estimates
derived in NTC00 using spectral types that were derived from narrowband
imaging, and which was in principle deep enough to probe the extinction 
seen to background stars.  From 106 matched sources
      \footnote{From this source matching, we find that the NTC00
       coordinates are systematically offset $1\farcm5$ south relative 
       to the positional grid defined by 2MASS.}
with non-zero extinction estimates in both catalogs, we find that they are fairly
similar, with $<\ak>_{NTC00}\,=\;0.47\,(0.38)$ compared to $<\ak>\,=\;0.35\,(0.26)$
for our work.
Direct comparison of the EPDFs constructed from these two studies confirmed that
they appear consistent with the NTC00 EPDF being somewhat broader.
The effective difference, ($\delta\ak\,\sim\,0.1$ between the $<\ak>$), between 
these EPDFs is much smaller than the bins of the $K$ band luminosity functions
we are employing in our subsequent mass function analysis.
Thus, it should not affect our derivation of and overall conclusions about 
the IMF of IC~348. 
On the other hand, it may bring added uncertainty to our statistical census of
very low mass brown dwarfs ($\mass\,<\,0.02\solarmass$; see also 
Section \ref{sec:discuss:imf_klf:substar}).
Since our study is better sampled (using more stars) and probes a much larger
volume of the cluster region than prior studies, we used the EPDFs derived
here when correcting for background field-stars (Section \ref{sec:lf:fieldstar})
and when modeling of the cluster's luminosity function (Section \ref{sec:imf:model}).


\placefigure{fig:jhklfs}

\section{The Infrared Luminosity Functions of IC~348}
\label{sec:lf}

\subsection{Constructing Infrared Luminosity Functions}
\label{sec:lf:rawlfs}

We constructed the raw infrared ($JHK$) luminosity functions (LFs)
for sources in the FLAMINGOS IC~348 region by using relatively
wide (0.5 mag) bins and by restricting the sources to those in
an area bounded by the $10.33\,\arcmin$ radius.
When we compared the $J$, $H$ and $K$ band IC~348 LFs, we found
that all of them display a double peaked structure: the first
peak lying at $J=H=K=\,13 \sim 13.5$ and the second between
$J$ = 17.5 and $K$ = 16.5.  Following our analysis of the
cluster's color-magnitude diagrams in Section \ref{sec:images:results},
we interpret the brighter of these peaks to be sources in the
IC~348 cluster and the fainter peak to be dominated by background
field-stars and galaxies. 

\placefigure{fig:klfs}

In Figure \ref{fig:klfs}, we show the $K$ band LFs of the IC~348,
breaking the cluster into the ``core'' and the ``halo'' sub-regions
and scaling them to stars per square degree. Further, we compare
them to the un-reddened field-star KLF constructed from our off-cluster
K band datasets. Confirming our preliminary study of the cluster's
structure from the radial profile, both sub-regions display a
considerable excess of sources relative to the field-star KLF for $\mk<15$.
Both sub-region KLFs reach bright peaks although they occur in somewhat
different locations with the core KLF, for example, peaking 1.5 magnitudes
brighter than the halo. Below these bright peaks, both KLFs flatten or
turnover before rising, parallel to the field-star KLF. 
While the faint KLF peak of the two sub-regions have nearly identical size
and structure, they appear to be smaller and shifted to fainter magnitudes
then the field-star KLF. Such differences could certainly be caused by the
reddening and obscuration of background field-stars due to the molecular cloud. 
Thus, we used our detailed study of the cluster reddening properties
(Section \ref{sec:images:redden}) to estimate the size of the
field-star contribution to the observed KLFs.

\subsection{Field-Star Correction to the Cluster KLF(s)}
\label{sec:lf:fieldstar}

To statistically estimate the field-star contribution to the raw IC~348 KLF(s), 
we used a Monte Carlo integration to convolve this control KLF with the effects
of a reddening probability distribution function that characterizes the 
the molecular cloud.  We treated the cluster's core and halo sub-regions
separately, and we used the extinction probability distribution functions
derived in Section \ref{sec:images:redden} from the fainter, likely
background stars de-reddened to the M dwarf locus.  In Figure(s)
\ref{fig:diffs}ab we compare the raw KLFs to the reddened field-star
KLFs appropriate to that sub-region, scaling the reddened field-star
KLF {\em only} by the ratio of the off-cluster area to the physical area
of the cluster or sub-region.

\placefigure{fig:diffs}

The reddened field-star KLFs very closely match the raw KLFs for $\mk>15$ in
both sub-regions, although they exceed the cluster KLFs at the faintest
magnitudes because they were not filtered to match our detection limits.
We subtracted these reddened field-star KLFs from the raw cluster KLFs, and
display the resulting differential KLFs in Figures \ref{fig:diffs}cd,
constructing error bars that are the $1\sigma$ counting statistics of the sum
of the cluster and reddened field star KLFs. The structure of the differential
KLFs is significant for the bins brighter than $\mk\,=\,16.75$, below which the
field-star correction clearly over-estimates the observations.
This magnitude limit corresponds to a 10\,\jupmass brown dwarf at the 2 Myr mean
age we assumed for IC~348.  Above this limit two sub-regions have a nearly identical
number of members, with the core containing $153\pm16$ sources with $\mk\leq15.25$
($172\pm24; \mk\leq16.75$) and the halo containing $150\pm23$ sources with
$\mk\leq15.25$ ($176\pm40; \mk\leq16.75$).  While the sub-region KLFs are well 
populated for $\mk<15$, at fainter magnitudes the substantial field-star 
corrections yield very large uncertainties in the KLF structure.

In Figure \ref{fig:cluslf} we directly compare the sub-region differential
KLFs and sum them to construct the overall cluster KLF. 
Again, the differences in the structure of the two sub-region KLFs are
obvious to the eye, with the peak of the halo KLF skewing significantly to
fainter magnitudes, before strongly turning over.  While the peaks of
the sub-region KLFs are clearly distinct, the KLFs are similar to within
their $1\sigma$ error bars in most bins but separated by $2\mbox{ to }3\,\sigma$
in the $\mk=11\mbox{ and }\mk=13$ bins.  A two sample chi-square
test of the two histograms (range: $\mk=8 - 15$) yielded a probability of
0.04 that they are drawn from the same parent distribution, indicating
that these sub-region KLFs are different at the $2\sigma$ level.

\placefigure{fig:cluslf}

When the two sub-region KLFs are summed together, the complete cluster KLF
has a very broad peak between $\mk=11.5-13$ before decreasing very
sharply to $\mk=15$.  For the combined cluster KLF
Despite the large uncertainties at faint magnitudes the composite KLF has
a statistically significant number of cluster members with $\mk>15$.  
Further, we found that if we increased the size of the field-star correction 
by using a bluer extinction distribution function (one relevant for the 
cluster stars), the main cluster KLF characteristics were not substantially 
altered, although the size of the very faint population is almost halved.


\section{The Initial Mass Function of IC~348}
\label{sec:imf}

To analyze the IC~348 differential $K$ band luminosity function(s) constructed
in Section \ref{sec:lf} we used our model luminosity function algorithm first
presented in \citet{mll00} and expanded in Paper I. Our goal was to place 
constraints on the initial mass function of IC~348 by deriving that mass
function or set of mass functions whose model luminosity functions best fit
the cluster KLF. For the purpose of comparing our work to other studies,
we individually analyzed the KLFs of the two cluster sub-regions as well as
the composite cluster KLF.
Since a young cluster's luminosity function is the product of an age
dependent mass-luminosity relation and the cluster's IMF,
we examined both the star forming history of IC~348 as derived by 
existing spectroscopic studies (Section \ref{sec:imf:sfh})
and the appropriate theoretical mass-luminosity relations
(Section \ref{sec:imf:dist}).
We then fixed these quantities and derived the cluster IMF
(Section \ref{sec:imf:model}).

\subsection{The Star Formation History of IC~348}
\label{sec:imf:sfh}

\placefigure{fig:sfh}

To derive accurate mass function for a young, partially embedded cluster,
our method requires a good estimate of the cluster's mean age \citep{mll00}.
We examined published spectroscopic studies of IC~348, which provide individual
age estimates for a reasonable large number of members.
In Figure \ref{fig:sfh}, we plot a histogram of the ages derived by
\citet{herbig98} using the de-reddened \vvi~color-magnitude diagram for
candidate pre-main sequence members in an 112 square arc-minute region of IC~348. 
We merged the ages of stars with and without detectable \halpha~emission and
derived an ensemble cluster mean age of $\sim2$ Myr.  We approximated the age spread
of IC~348 to be $\sim3$ Myr, corresponding to constant star formation from
0.5 to 3.5 Myr ago, and we note that this age spread is 2.5 times longer
than that we approximated for the younger Trapezium Cluster. 
We show in Figure \ref{fig:sfh} that our assumed star forming history closely
approximates the bulk of the \citeauthor{herbig98} SFH, but clips the
``older'' tail of this distribution. 

We did not include this ``older'' tail in our SFH of IC~348 for the following reasons: 
1) Observational uncertainties (for example, membership criteria or errors
in photometric and spectral classification) in the derivation of de-reddened
color-magnitude or theoretical HR diagrams lead to exaggerated star forming 
histories, specifically resulting in artificially inflated cluster's age spreads
\citep{hart01,kh90};
2) A number of the ``oldest'' IC~348 objects analyzed in the NTC00 NICMOS study
were also found to be the reddest cluster objects, suggesting to us that they
are background interlopers;
3) Our assumed age spread very closely approximates that SFH derived by
\citet{ps00}, who found that the star formation in IC~348 began approximately 3 Myr
ago with negligible star formation at older ages, and finally;
4) Model luminosity functions are very insensitive to the assumed age
spread \citep{mll00}.
Further, \citeauthor{ps00} found no dependence of the SFH on location within IC~348. 
Thus, we used the same star formation history when modeling both
the ensemble cluster and the cluster's ``core'' and ``halo'' sub-region KLFs.

\subsection{The Cluster Distance and the Mass-Luminosity Relation}
\label{sec:imf:dist}

\citet{herbig98} included an extensive discussion on the distance to IC~348 
based upon literature sources existing at that time.  Arguing that closer 
distances ($\sim260$~pc) were systematic under-estimates, he chose a distance 
for IC~348 (320 pc) based in part upon the fact that within the current 
uncertainties, one could not differentiate between the distance to IC~348
\citep[$316\,\pm 22 $~pc][]{strom74} and to the Perseus OB2 association
\citet[$322\,\pm 30 $~pc][]{borg64}.  
Recent studies of this regions using Hipparcos data do not appear to have
resolved the distance uncertainty between the Perseus OB2 association and 
the IC~348 cluster.  \citet[][]{hippo99} derived a distance of $317 \pm 27$~pc 
for 17 members of the Perseus OB2 association spread over a projected
$37\,\times\,37$ pc area, while also statistically estimating the number
of interlopers. A study by \citet{belikov02}, including a larger subset of
probable members, seems to confirm this distance to the OB association.
On the other hand, when \citet{scholz99} performed a recent proper motion
study of the IC~348 region, they derived a distance of $\sim 261$~pc using
a set of 9 Hipparcos sources.  
This latter distance estimate to IC~348 should probably be treated 
with some caution.  Since more than half of these 9 sources with 
parallaxes were at projected distances of 2.5 - 8 pc from the cluster center,
they fall well outside any cluster outer radius we have discussed here and may
not be actual members, especially since no statistical estimate of the number
of non-members was performed for the \citeauthor{scholz99} study. 
Further, \citet{ripepi02} very recently reported the discovery of an
F star within the IC~348 boundaries that displays rapid $\delta$
Scuti-like variability that is interpreted as the pulsation of
a PMS star while in its instability strip \citep{marconi98}.
The derived pulsation period strongly favors a distance of $\sim320$ to IC~348.
Thus, it remains unclear if the distance to IC~348 can be separated from the
distance to the OB association.  For our modeling we adopted the distance
of 320 pc (m-M = 7.5) to IC~348 for consistency with the work of LL95 and
\citet{herbig98}, but we examined how such a distance uncertainty could affect
our results.

\placefigure{fig:kml}

Such distance uncertainties may systematically affect the derivation of cluster
properties such as mean age and possibly the IMF.  
\citeauthor{herbig98}, for example, showed that assigning IC~348 a closer 
distance yielded a systematically older cluster mean age since the cluster 
appears intrinsically fainter when compared to pre-main sequence evolutionary 
models.  In Figure \ref{fig:kml}, we examine the net effect of this age-distance
uncertainty on the theoretical mass-luminosity relations relevant for our 
luminosity function modeling.  Shifting the distance from 320 to 
260pc produces a shift in the mean age from 2 to $\sim3-4$ Myr \citep[see 
also][]{hll01}. However, the change in the distance modulus ($-0.42$ magnitudes)
is roughly equivalent to the average luminosity evolution of stars (1.0 - 0.1
\solarmass) between 2 and 3 Myr ($dK\sim\;-0.35$).  By comparing the mass-$K$
magnitude relation at 2 Myr and 320pc to that at 3 Myr and 260 pc, we find
that above the hydrogen burning limit, our derived IMF will have little
systematic dependence upon the cited age-distance uncertainty for IC~348 and
should be a faithful representation of the true cluster IMF.  On the other
hand, the slope of the substellar mass-$K$ magnitude relation is systematically
affected, in part due to a larger mass range undergoing deuterium burning.
Thus, our derived substellar IMF will be less reliable than the stellar IMF
until this age-distance uncertainty is resolved.

Lastly, in Figure \ref{fig:kml} we also compare the theoretical mass-luminosity 
relations taken from two sets of evolutionary calculations. As we found in 
Paper I, different theoretical mass-$K$ magnitude relations are very consistent
between current sets of PMS tracks, meaning that the derived IMF will be mostly
independent of which set of modern PMS models we use.  To provide consistency
between our studies of various young clusters, we derived the cluster
IMF using our standard set of PMS tracks, which are primarily based on the
\citet{dm97} tracks and whose construction and conversion to observable quantities
were described in \citet{mll00}.

\subsection{Other Modeling Parameters: Reddening and Binaries}
\label{sec:imf:red}

Three additional characteristics of young stars that we can include into our 
model luminosity function algorithm are the reddening of the cluster members
by the parental molecular cloud, excess infrared flux due to optically 
thick disks around the cluster members, and the frequency of un-resolved 
companions.  First, to account for the reddening of the cluster by the Perseus
Molecular Cloud, we used the extinction distribution functions derived in
section \ref{sec:images:redden} for the bright cluster stars de-reddened
to the cTTS locus.

Second, in Section \ref{sec:images:results} we confirmed the LL95 
finding of $\sim60$ infrared excess sources distributed across the
IC~348 cluster region. Correcting for field stars and dividing the
excess sources by sub-region, we found infrared excess fractions
of~$14\pm11\%\;(22/153)$~for the core sub-region, ~$24\pm17\%\;(36/150)$~for
the cluster halo and~$19\pm10\%\;(58/303)$ for the composite cluster.
If we restrict these estimates to those 42 sources which have NIR excess
greater than their $1\sigma$ photometric errors, these percentages drop
to $9\%, 14\%\mbox{ and }11\%$, respectively.
Because these excess fractions are very small, we chose to no account
for NIR excess in our KLF models of IC~348.
It is possible, however, that a more substantial fraction of sources 
could have small $K$ band excesses ($dK_{irx}\sim0.1$) that are not
apparent in the color-color diagram especially since \citet{hll01} found
that $65\%$ of the IC~348 members  have inner circumstellar disks as
traced at $3\micron$.  We argue such small excesses
will not substantially affect the cluster KLF and hence the derived
cluster IMF since this flux excess is much smaller than the bins used
to create the cluster luminosity function.

Finally, we choose not to include un-resolved binaries into our modeling of
IC~348, paralleling our analysis of the Trapezium Cluster in Paper I. 
Thus, the IC~348 IMF we derive is the ``primary'' or ``single'' star IMF. 
\citet{dbs99} found that the binary fraction in IC~348 is very similar to
that found in other young clusters and to the field. 
Further, the consistency of the binary frequency in stellar clusters over
a large range of cluster age \citep{pat02}, including young clusters like
IC~348, suggests that the single star IMFs of star clusters can be readily
derived and compared without correction for binaries.

A related issue in our studies of different cluster IMFs is whether or not a 
significant fraction of {\it resolved} wide binaries have been included 
into the cluster KLFs. For example, the physical resolution of our Trapezium 
study was $\sim240$ au, while in IC~348 it is $\sim480$ au.  Since the binary 
fraction at these large separations is quite small ($f<0.1$), the few resolved 
binary systems that will be included into these cluster KLFs should not
significantly modify the single star IMFs we derive.

\subsection{Modeling the IC~348 Differential KLF(s)}
\label{sec:imf:model}

Using these cluster parameters, we produced a suite of model luminosity
functions by varying the underlying initial mass functions. Our goal is
to find the simplest underlying IMF or range of IMFs whose model luminosity
functions best fit the observed cluster KLF(s).  Our standard IMF parameterization
consists of power-law segments, $\Gamma_{i}$, connected at break masses,
$m_{j}$, and we use the fewest segments necessary to fit the model KLF,
which for IC~348 resulted in 2 and 3 segment IMFs.
We independently calculated and fit model luminosity functions for both
the composite cluster KLF and the cluster's ``core'' and  ``halo'' sub-region KLFs,
varying the extinction distribution function appropriately
for each region.  Our fitting technique calculates the \chisq~statistic 
and probability between the model KLFs and observed KLF over a range of 
magnitude bins and from these statistical measures, the parameters (mean, 
standard deviation) of the underlying IMF are derived.  We summarize these 
fits and the resulting IMFs for the complete cluster in Section \ref{sec:imf:model:clus},
and for the two sub-regions in Section \ref{sec:imf:model:corehalo}

\placetable{tab:imf:fits}

\placefigure{fig:klf_imf_clus}

\subsubsection{The Composite IC~348 KLF}
\label{sec:imf:model:clus}

We examine the best fit model KLFs to the complete cluster KLF in the left hand panel 
of Figure \ref{fig:klf_imf_clus}.  The composite cluster KLF displays a very broad main 
peak followed by a sharp turnover at fainter magnitudes. This structure required the 
use of model KLFs based upon a 3 power-law underlying IMF, and we tabulate the derived 
IMF parameters in Table \ref{tab:imf:fits}.  By varying the range of KLF bins fit by 
our $\chisq$~routine, we find that with these 3 segment underlying IMFs we were  
able to fit the composite cluster KLF down to $\mk=15$, constraining the cluster IMF 
from $2.5\solarmass$ down to $\sim0.035\jupmass$ (for $<\av>=4$).
Non-power-law structure in the cluster KLF for $\mk>15$ prevented decent fits to
these bins (see Section \ref{sec:discuss:imf_klf:substar}).

Over this $\mk-\mass$ fit range, our KLF fits yielded an IC~348 IMF having a broad peak
down to the hydrogen burning limit before rolling over and decreasing sharply into the 
substellar regime.  The high mass $\Gamma_{1}$ slope was moderately constrained with 
an index of $\sim-1.5\pm0.3\; (\mbox{Salpeter }=-1.35)$ before flattening at
$m_{1}\;=\;0.80\solarmass$.  The $\Gamma_{2}$ slope and the $m_{2}$ mass break are 
both strongly constrained by the broad KLF peak and the sharp KLF turnover at $\mk=13$, 
with a very slowly rising  $\Gamma_{2}\,\sim-0.2\pm0.15$ across much of the IMF before 
peaking at the hydrogen burning limit.  Because of the rapid change in KLF slope 
between $\mk=12-15$, there is a moderately large uncertainty in $\Gamma_{3}$ IMF 
parameter with a sub-stellar IMF slope that is very steeply falling with an index 
of $\sim+2.0\pm0.4$.

\subsubsection{The IC~348 Core and Halo Sub-Region KLFs}
\label{sec:imf:model:corehalo}

In the left hand panels of Figure \ref{fig:klf_imf_rad} we display the sub-region
differential KLFs compared to the best fit model KLFs derived from our \chisq~fitting
technique.  The underlying power-law IMFs are displayed in the right hand panels of
this figure, and we list our derived IMF parameters in Table \ref{tab:imf:fits} as 
a function of fit range.

\placefigure{fig:klf_imf_rad}

Unlike the more complex structure of the composite cluster KLF, we found that model 
KLFs constructed using 2 power-law IMFs provided satisfactory fits to both sub-region
KLFs, although we were able to obtain slightly better fits to the ``halo'' KLF using a
3 segment IMF.  For both sub-regions, the $m_{1}$ mass break was very strongly
constrained by the location of the bright KLF peaks in the 2 power-law IMF fits. 
The $\Gamma_{1}$ IMF slope rises steeply with decreasing mass in the ``'core'' 
sub-region, but is much shallower in the ``halo'' region. 
In our 3 power-law fits to the ``halo'' sub-region, we found a modest
$\chisq$~minima that indicated there is an inflection in this high mass slope
near $0.80\solarmass$, yielding a $\Gamma_{1}$ similar to that found in the
2 segment ``core'' IMF and to the composite cluster IMF
(see Table \ref{tab:imf:fits} for comparison of 2 and 3 segment IMFs for IC~348).
As evident in their KLFs, the two sub-regions have distinctly different IMF
peaks regardless of the complexity of the underlying IMF, with the
core $m_{1}$ mass break occurring at $0.56\,\pm0.18\,\solarmass$ and the peak
of the halo IMF strongly constrained to lie at $0.10\,\pm0.02\,\solarmass$.  
Below their peaks, both sub-region IMFs steadily fall with decreasing mass down
to the $\mk-\mass$ fit limit, which for both sub-regions corresponded to at
least the $\mk=15$ bin and a minimum fit mass of $0.035\,\solarmass$. 
The $\Gamma_{2}$ slope was more tightly constrained for the two power-law
``core'' IMF than for the halo, and we found that a single power-law IMF with slope of
$\Gamma\,\sim\,0.4$ could fit the core KLF from the IMF peak ($\sim0.5\,\solarmass$) all 
the way down to the $\mk=16.5$ bin, which has a lower $\mk-\mass$~limit from the DM97 
tracks of $0.017\,\jupmass$.  Similar to the composite cluster IMF, the rapid change in 
halo KLF slope between $\mk=12-15$ yielded considerable uncertainty in the corresponding 
IMF parameter below the halo IMF peak, but it was constrained to be steeply falling,
with a $\Gamma_{bds, halo}\,\sim\,2.0$.  Despite any limitations to our fitting
procedure, we found that while the resulting substellar IMF slopes are different
in the two sub-regions
($\Gamma_{bds, halo}\,\sim\,2.0$ versus $\Gamma_{bds, core}\,\sim\,0.4$), 
they correspond to similar proportions of brown dwarfs between the two regions,
as we will discuss further in Section \ref{sec:discuss:radimf}.


\section{Discussion}
\label{sec:discuss}

\subsection{The KLFs and IMFs of IC~348 and the Trapezium}
\label{sec:discuss:imf_klf}

\placefigure{fig:trap_ic348}

\subsubsection{The Stellar Regime}
\label{sec:discuss:imf_klf:stellar}

We examined the structure of the IC~348 KLF and derived IMF by comparing these
characteristics to those we derived for the Trapezium cluster in Paper I. 
By comparing the Paper I Trapezium IMF (also reproduced in Table \ref{tab:imf:fits})
to the derived IC~348 IMF we find that these two very young clusters have nearly
identical IMFs throughout the stellar regime. 
As illustrated in Figure \ref{fig:klf_imf_clus}, both clusters
have IMFs that rise in number with decreasing mass into the sub-solar
regime, with Salpeter-like power-law slopes ($\Gamma_{1,\;Trap}\,=\,-1.21$
and $\Gamma_{1,\;IC\,348}\,=\;-1.53$).  Their IMFs both flatten around
$0.7 \solarmass$, having slopes of $\Gamma_{2}\,\sim\,-0.2$ and form 
very broad shallow peaks at sub-solar masses. The ``peak'' or mode of 
their IMFs varies between $0.15\mbox{ and }0.08 \solarmass$ with a skewing
of the mode of the IC~348 IMF to slightly lower masses than the Trapezium IMF,
although this may be due to the Trapezium being only core of the larger Orion
Nebula Cluster (see Section \ref{sec:discuss:radimf}).

The strong similarities between these clusters' stellar IMFs exist despite
the significant apparent differences between their cluster KLFs.  
In Figure \ref{fig:trap_ic348}a, we compare the Trapezium and IC~348 KLFs,
shifting them by their respective distance moduli to absolute magnitudes
and scaling the size of the Trapezium population to match that of IC~348. 
The IC~348 KLF is clearly broader and shifted to fainter magnitudes relative to
the Trapezium KLF with their primary KLF peaks differing by almost 2 magnitudes.
These differences, however, are precisely those predicted by the evolution
of the luminosity function with age \citep{ll95, mll00}. Indeed if we evolve
a model KLF of the younger Trapezium to the age of IC~348 and compare it
to the IC~348 KLF in Figure \ref{fig:trap_ic348}b, we find it agrees
with the observed IC~348 KLF extremely well down to $\mk=12.5$, near 
the unreddened hydrogen burning limit for IC~348
($\tau=2$~Myr, ${\mk}_{,\, HBL}\,\sim\,12.7$)
\footnote{This quantity is fairly independent of current PMS tracks,
see Figure \ref{fig:kml}. From various models we find for IC~348 that~
${\mk}_{,\, HBL}\mbox{\citep{dm94}}\,=\,12.74$,~
${\mk}_{,\, HBL}\mbox{(DM97)}\,=\,12.67$,~
${\mk}_{,\, HBL}\mbox{\citep{bur97}}\,=\,12.55$,~and~
${\mk}_{,\, HBL}\mbox{\citep{bcah98}}\,=\,12.83$. 
For a mean age of 3 Myr at a distance of 260 pc,
$<{\mk}_{,\, HBL}>\,=\,12.68$. }.
Fainter than this magnitude the evolved Trapezium KLF moderately 
under-estimates the IC~348 KLF between $\mk=13-13.5$, indicative 
of the slight skewing of the mode of the derived IC~348 IMF to
lower masses.  Below $\mk=13$, both the observed IC~348 KLF and the evolved 
Trapezium KLF steeply fall in number, marking the transition 
to the substellar regime that we explore in the next section.

\subsubsection{The Sub-Stellar Regime} 
\label{sec:discuss:imf_klf:substar}

Reviewing Figure(s) \ref{fig:klf_imf_clus}b and \ref{fig:trap_ic348}, 
one immediate similarity between the substellar KLFs and IMFs of IC~348 
and the Trapezium is their mutual steep decline towards fainter magnitudes 
and lower masses.  Although the IC~348 IMF mode skews to lower masses
than the Trapezium, it turns over and decreases in a much steeper 
manner than the Trapezium, i.e., 
$\Gamma_{BDs,\,IC348}\,=\,2.0$ while $\Gamma_{BDs,\,Trap}\,\sim\,0.7$.
This is also illustrated by the way that the evolved Trapezium KLF 
seems to consistently over-estimate the number faint objects in IC~348,
although they agree within the large statistical uncertainties. 

A second similarity between the substellar KLFs of IC~348 and the Trapezium,
is the formation of a modest secondary KLF peak.
After correction for background stars we estimate that in IC~348
this secondary KLF peak contains $42\pm29$ sources between $\mk=15-17$. 
Further, the secondary KLF peak for IC~348 occurs precisely in the magnitude
range predicted by the evolved Trapezium KLF (see Figure \ref{fig:klf_imf_clus}b),
suggesting they are related features and correspond to sources in the mass
range from $10 - 20 \jupmass$.
As was the case for our Trapezium modeling, such KLF structure rejected
our fitting of these magnitude bins using models based upon 3 segment
power-law IMFs.  This is because we are trying to fit this non-power law KLF
structure (a dip or gap followed by a secondary peak) with model KLFs that
are essentially a power-law throughout the brown dwarf regime because the
theoretical mass-luminosity relations that we have used are smooth and
do not contain any significant evolutionary features in this mass range.
Considering the statistical uncertainties due to the background correction
we did not attempt to explicitly fit additional IMF segments to the IC~348
peak although the size of the predicted peak using the Trapezium IMF 
closely approximates this feature.  We discuss the origin of this
secondary KLF peak further in Section \ref{sec:discuss:mlr}.

One meaningful constraint regardless of detailed IMF structure is the 
{\em fraction} of cluster members that are substellar.  For example,
in the Trapezium we found that $22_{\,-2}^{\,+4}\,\%$ of the clusters members
fell between $80\mbox{ and }17 \jupmass$, having the interesting implication that
only $\sim\mathit{1 \mbox{ in } 4}$ of the cluster members were brown dwarfs! 
Estimating this fraction for IC~348 was more difficult because the background
correction makes directly counting all the substellar sources difficult.
Rather than directly counting sources, we first integrated the derived IMF to
calculate the fraction of substellar sources down to the mass limit of our fit.
We found that brown dwarfs between 80 and 25 \jupmass~(the lower mass limit for
$\av=0$ and corresponding to our model fits to $\mk=15$ and the beginning of
the secondary KLF peak) constitute only $14\%$ of the members with $2\%$
uncertainty due to the variation in our fits, and $9\%$ uncertainty due to
the counting statistics which are dominated by the size of the background correction. 
For comparison, integrating the Trapezium IMF over the same mass range yields
a brown dwarf fraction of $20\%$.

When we count the number of sources contained in the secondary peak of the
IC~348 KLF and not included into the KLF fits, we were able to extend the
range of substellar masses down to $10\,\jupmass$, however we also found
that the precision of our estimate is significantly worsened by the large
field star contamination. Since this secondary KLF peak contains between
26 and 42 sources depending upon the size of the correction for field stars,
the total substellar fraction increased to $20-25\%$ for IC~348, although
the corresponding error bar also increased to $\pm14\%$. Thus, despite the
uncertainties due to the field star correction these fractions indicate
that IC~348 has a very similar brown dwarf fraction to that found for
the Trapezium and further, that brown dwarfs are not nearly
as populous as stars in either cluster.

\subsection{The Secondary Sub-Stellar Peak in the Cluster KLFs}
\label{sec:discuss:mlr}

As we derived in Paper I, the secondary peak in the Trapezium KLF can be
attributed to a corresponding IMF peak near the deuterium burning limit. 
Our finding of an IC~348 KLF feature having a similar size and corresponding
to the same mass range, $10 - 20\,\jupmass$, lends some support to this
conclusion. One interesting implication of such structure in the cluster's
substellar IMF would be the existance of a separate formation mechanism for
very low mass brown dwarfs, as we postulated in Paper I. However, since the LF
is the product of the IMF and the slope of the mass-luminosity relation, 
it may also be the case that a subtle feature or inflection exists in the
empirical mass-luminosity relation that has not been resolved by even the
most recent theoretical models of brown dwarfs \citep[e.g., ][]{cbah00, bcah02}. 
Such an unaccounted for M-L feature could also produce features in the LF
independent of the structure of the underlying IMF, as previously found,
for example, in the \citet{ktg90, ktg93} studies of the $\mbox{H}^{-}$ and 
$\mbox{H}_{2}$ opacity features in the optical $\mass-\Mv$ relation for field stars.

There is very recent observational evidence that seems to point toward an
unaccounted for M-L feature for brown dwarfs. \citet{jam02} discuss a feature
in the color-magnitude diagram for the Pleiades Cluster, corresponding
to an apparent gap in both color and luminosity space for substellar objects.
\citet{jam02} draw an analogy between this feature and the $\mbox{H}^{\,-}$
opacity related Wielen gap \citep{ktg90}, attempt to correct the
theoretical mass-luminosity relation for the effect of such a gap, and
conclude that the lack of this feature in the theoretical tracks
(which they suggest is related to the formation of dust in the brown dwarf atmospheres)
is likely the cause of the IMF structure we derived for the Trapezium in Paper I
and that was also derived for the IMF of the $\sigma$-Orionis cluster by
\citet{bejar01}. We point out that what we observe in the Trapezium and
IC~348 appear to be secondary peaks in the cluster KLFs.
Drawing an analogy to studies of the $\mbox{H}_{\,2}$ opacity feature in
the field star mass-luminosity relation \citep{ktg93,kt97} or to the
effects of deuterium burning on theoretical model KLFs \citep{zin93, mll00}
such a peak could be caused by a piling up of stars with a range of mass in
a narrow range of luminosity. If we were to repeat the \citet{jam02} correction
to the theoretical M-L relation but assume a more steeply falling slope for the
substellar IMF of the Pleiades, such an inflection or piling up would be the
result.

Regardless, it is uncertain which effect (IMF, M-L relation) will in fact
dominate the nature of the secondary KLF peaks we have observed. 
Clearly, improvements are needed in the theoretical tracks, which,
in combination with future observations, should be able to disentangle
these two effects for the substellar IMFs of young clusters.
However,  which mechanism is producing this KLF peak appears to operate only
at the very lowest masses in the very young ($\tau<10$ Myr) clusters we are 
studying. Thus, it seems unlikely that it will modify the derived IMF at the
hydrogen burning limit and future updates probably will not adjust (inflate or
decrease) the percentage of sources that are brown dwarfs in these clusters.

\subsection{Radial Variation of the IC~348 IMF}
\label{sec:discuss:radimf}

Our division of the cluster into two sub-regions based on the cluster's radial
profile (see Section \ref{sec:images:radprof}) allowed us to make important
comparisons to past studies of the IC~348 IMF, which have primarily focused
on the cluster's core. For example, when we compare in Figure \ref{fig:radimf}
the IMF we derived for around 150 sources in the IC~348 ``core'' to those 
MFs derived by NTC00 and \citet{luh2000} each containing 90-100 sources,
we find they agree remarkably well, all having an IMF that peaks or has a mode
between 0.3 and 0.5 \solarmass and turns over into the substellar regime.
Those differences that do exist between our ``core'' IMF and these studies are
likely due to the $50\%$ difference in sample sizes, which may produce statistical
flucuations, or source saturation in the NICMOS study, which causes it to
exclude higher mass stars.

\placefigure{fig:radimf}

Such a strong similarity is not found, however, when we compare the IMF of the cluster's
core (derived in this or past studies) to that we derived for the cluster halo.
Paralleling the apparent physical differences in the KLFs of the IC~348
sub-regions (see Figures \ref{fig:klfs} and \ref{fig:cluslf}),
we find that the IMF of the halo skews to lower masses relative to the 
IMF of the cluster's core.  This variation of the IMF is a real representation
of the differences in the two sub-regions and is not the product of a
variation in some other physical quantity.  Reddening due to the 
molecular cloud and specific to each sub-region is included into these fits,
although no meaningful differences exist between the relevant EPDFs 
of the two regions (see Section \ref{sec:images:epdfs}). Although we did not
include the effects of infrared excess, any such effect would actually {\em increase}
the IMF differences, since the larger excess fraction of the halo and would
require the halo IMF peak to shift to lower masses.  Further, the larger excess
fraction in the halo might imply that the halo is younger than the core.
However, accounting for such an age difference would again shift the IMF peaks
in {\em opposite} directions from one another and amplify the IMF variations we derive.
Lastly, although it would seem to contradict the distribution of infrared excess
and $\halpha$~sources and was not corroborated by the study of \citet{ps00},
\citet{herbig98} reported a slight age gradient in IC~348, finding an increasing
mean age at larger radii. Such a gradient does act in the correct direction to
account for some of the KLF differences, however, even if this age gradient is real,
it is considerably too small ($\tau\,\sim\,1.45$ Myr at $R=4\,\arcmin$ to
2.8 Myr at $R=10\,\arcmin$) to bring the IMFs of the sub-regions into agreement.
To align the sub-region IMF peaks given the observed KLFs would require two distinct
populations where the halo is 5-10 Myr older than the core. This is an difficult
hypothesis to accept considering the populations are of equal size, and while such
a model might align the peak of the IMFs by shifting the halo to higher masses,
it would further imply that the IMF of the halo is truncated below
$0.3\,\solarmass$.  As was originally found in the model fits of LL95,
we conclude that such a two age population model cannot explain the IC~348 KLFs.
Indeed, the one physical phenomena that we cannot test here, the fact that the
``core'' sub-region contains ``halo'' members due to the two dimensional projection
of the cluster, implies a priori that the spatial IMF variation are {\em larger}
than we derive since the ``halo'' IMF contaminates the cluster ``core''(but not the
reverse).

Since differences between the core and halo sub-region KLFs can only
be related to their underlying IMFs and these IMF differences (as illustrated 
in Figure \ref{fig:radimf}) appear statistically significant at the $2\,\sigma$
level, we conclude that a physical variation in the IC~348 IMF appears to exist
on spatial scales of the order of 1 parsec. 
Further, these radial IMF differences primarily occur in a limited range
of sub-solar masses between 0.5 and 0.08 \solarmass in IC~348.
This radial IMF variation is unlike typical scenarios for mass segregation
in very young clusters in which only the higher mass stars ($\mass>1\solarmass$)
are thought to be preferentially affected. With these cluster(s) being only
a few crossing times old, there is thought to be only enough time for the massive
stars to sink to the cluster core \citep{kah00}; otherwise, these more massive stars
are thought to be preferentially born in the cluster center through a
process of competitive accretion \citep{bonn01}.

Radial variations of a cluster's sub-solar IMF have also been reported for
the Orion Nebula Cluster(ONC), of which the Trapezium is the core. While
this cluster displays additional evidence for the segregation of high mass
stars to the cluster core \citep{hil98a}, the low mass ONC IMF varies between
the central core and outer cluster halo.  This has been shown by \citet{lah97}
and \citet{hc2000} who found that while the IMF of the central
$\mbox{r}<0.35\,\mbox{pc}$ Trapezium core peaks around $0.2\solarmass$, the addition
of the cluster's halo ($\mbox{r}_{tidal}=2.5\,\mbox{pc}$) produces a composite ONC
IMF that continues to rise down to $0.1\solarmass$, the completeness limit of the
\citeauthor{lah97} spectroscopic survey of the entire ONC cluster. 
Thus, this skewing of the stellar IMF with radius proceeds in the same direction
in the ONC as we find in IC~348, although IC~348 may in fact be a better location for
studying such mass segregation. This is because the ONC is projected along the
same line of sight as the Orion Ic OB association, which may contaminate the
derived cluster IMF at large radii.
What is clear in both clusters is that meaningful comparisons of their IMFs can
only be performed after composite IMFs have been derived over large physical areas
of the clusters; smaller areas are subject to mass segregation effects even at
very young ages.  Further, such effects must be taken into account when placing
them into discussion of a ``universal'' IMF \citep[see also][]{kroupasci02}.

Lastly, it is important to examine how the substellar sources are radially
distributed.  Since dynamical effects that could produce the radial IMF
variation may operate on the lowest mass sources by skewing them to larger
radii, brown dwarfs may correlate with the distribution of low mass,
$0.1-0.3 \solarmass$ sources and be systematically located further from the
cluster center.  While the slope of the substellar IMF varies significantly
between the IC~348 core and halo, the percentage of sources that are
substellar does not.
We found that sources in the mass range from $80 - 25 \jupmass$ constitute
roughly $14\%$ of the sources in the cluster core and a similar $16\%$ in
the cluster halo (poissonian error for both fractions is $\sim\,10\%$). 
This finding suggests that the mechanism that is breaking the universality of
the IMF on small spatial scales, whether it is primordial IMF variations or
dynamical mass segregation, does not appear to be significantly acting
upon the substellar population. Wide-field infrared imaging of other young
clusters, such as, for example, a survey of the substellar population in the
ONC halo, will be necessary to determine if similar, uniform spatial distributions
of brown dwarfs exist in other clusters that are still embedded in their 
parental molecular clouds.

\section{Conclusions}
\label{sec:conclude}

Using wide-field near-infrared images provided by the FLAMINGOS camera on the
Kitt Peak 2.1m telescope, we performed a detailed census of the young 2 Myr
IC~348 cluster located on the northeastern end of the Perseus Molecular Cloud.
Using the multi-color infrared photometry provided by our observations, we
explored this cluster's structure, reddening, and relationship to the parental
molecular cloud, and then used these results to construct and to analyze the
IC~348 KLF and to correct it for field star contamination.  Using our model
luminosity function algorithm described in \citet{mll00} and \citet{aam02},
we derived the cluster's initial mass function, providing detailed fits and
error estimates.  From our analysis of the cluster's structure and luminosity
function and by comparison to our earlier study of the Trapezium cluster, 
we draw the following conclusions about the KLF and IMF of IC~348:

\begin{enumerate}

\item We derive an IMF for the composite IC~348 cluster spanning the mass range
  from $2.5\mbox{ to }0.035\solarmass$. Further, we find that the IC~348 IMF 
  we derive is nearly identical to the the Trapezium IMF we derived in Paper I:
  The two clusters IMFs rise with decreasing mass, having a Salpeter-like slopes
  before flattening below $0.7\,\solarmass$.
  Within the current uncertainties in PMS evolutionary models, we find that
  the mode of these star cluster IMFs appear to fall at a
  mass of $\sim0.1-0.2\,\solarmass$.

\item Further, we find that the relative size of the substellar population
  is very similar in both clusters within the uncertainties of our method,
  revealing that brown dwarfs constitute between $15-25\%$ of the members of
  either cluster and do not dominate the cluster's stellar populations.
  
\item IC~348 forms a modest but significant secondary KLF peak, corresponding to
  sources in the same mass range that we found responsible for a similar
  secondary KLF peak in the Trapezium.  The similar KLF features in the
  Trapezium and IC~348 may signify either the presence of a secondary
  peak in the substellar IMF between $10-20\jupmass$~as we derived in Paper I,
  or be the result of a previously unknown feature in the brown dwarf
  mass-luminosity relation \citep[e.g., ][]{jam02}.

\item Radial variations are found in the KLF and IMF of IC~348 on the parsec
  scale, with a skewing of the KLF to fainter sources and the IMF to lower
  mass stars in the cluster's halo, a portion of the cluster whose IMF was
  previously undetermined. This radial variation in the sub-solar IMF is
  similar to but more pronounced than what was found previously for the sub-solar
  mass stars in the Orion Nebula Cluster. It is unclear what process is breaking
  the universality of the cluster's IMF on small spatial scales, but it appears
  different from dynamical mass segregation which primarily acts upon higher mass
  stars at these young ages. Further, while the slope of the substellar
  IMF varies as a function of radius in IC~348, the spatial distribution of 
  brown dwarfs in this cluster is different from the stars and is uniformly
  distributed between the cluster's core and halo.

\end{enumerate}

Finally, we draw the general conclusion that the existence of radial
variations of the IMF on parsec scales even at very young ages (1 Myr)
may mean that wide-field imaging surveys are a pre-requisite to making
meaningful IMF comparisons between different embedded clusters, a fact
that has clearly been true in older open clusters for some time.

\acknowledgments

We would like to thank Simon Hodgkin for a disscussion on the existence 
of a previously unknown feature in the substellar mass-luminosity relation.
EAL and AAM acknowledge support from a Research Corporation Innovation Award
and Presidential Early Career Award for Scientists and Engineers (NSF AST
9733367) to the University of Florida. The FLAMINGOS camera was designed
and constructed at the University of Florida under a grant from the National
Science Foundataion (NSF AST 9731180).
The data for this paper were collected under the NOAO Survey program,
``Toward a Complete Near-Infrared Spectroscopic and Imaging Survey of Giant
Molecular Clouds'' (E. Lada, P.I.).  This publication makes
use of data products from the Two Micron All Sky Survey, which is a joint project 
of the University of Massachusetts and the Infrared Processing and Analysis
Center/California Institute of Technology, funded by the National Aeronautics
and Space Administration and the National Science Foundation.

\clearpage




\clearpage

\begin{deluxetable}{llccccrcccr}
\tablecolumns{11}
\tablewidth{0pt}

\tabletypesize{\small}
\tablecaption{Summary of FLAMINGOS Observations of IC~348
\label{tab:obs}
}

\tablehead{
\colhead{Date}         &
\colhead{Julian Date}  &
\colhead{Target}       &
\colhead{Filter}       &
\colhead{Exp. }        &
\colhead{Dithers}      &
\colhead{Total Exp.}   &
\colhead{Airmass}      &
\colhead{Seeing}       &
\colhead{$5\sigma$}    &
\colhead{Flat-Field}   \\
\colhead{      }       &
\colhead{      }       &
\colhead{      }       &
\colhead{      }       &
\colhead{(sec) }       &
\colhead{      }       &
\colhead{(sec) }       &
\colhead{      }       &
\colhead{($\,\arcsec\,$)}       &
\colhead{(mag) }       &
\colhead{type  }
}

\startdata
14 Dec 2001 & 2452257.82961 & IC 348 & $K$   & 20 & 49 &  980 & 1.139 & 1.60 & 16.81 & dome \\
14 Dec 2001 & 2452257.85944 & IC 348 & $H$   & 20 & 48 &  960 & 1.262 & 1.81 & 17.69 & sky  \\
14 Dec 2001 & 2452257.88446 & IC 348 & $J$   & 20 & 46 &  920 & 1.414 & 1.87 & 17.70 & sky  \\
17 Dec 2001 & 2452260.81462 & Off 1  & $K$   & 30 & 30 &  900 & 1.112 & 2.10 & 16.97 & dome \\
08 Feb 2002 & 2452313.60532 & IC 348 & $\Ks$ & 60 & 16 &  960 & 1.007 & 1.48 & 17.72 & dome \\
08 Feb 2002 & 2452313.62027 & IC 348 & $H$   & 60 & 14 &  840 & 1.020 & 1.49 & 18.04 & sky  \\
08 Feb 2002 & 2452313.68697 & IC 348 & $J$   & 60 & 24 & 1440 & 1.176 & 1.67 & 18.82 & sky  \\
11 Feb 2002 & 2452316.58424 & Off 2  & $\Ks$ & 60 & 16 &  960 & 1.001 & 1.59 & 17.74 & dome
\enddata

\end{deluxetable}


\clearpage

\begin{deluxetable}{clccccc}
\tablecolumns{7}
\tablewidth{0pt}

\tabletypesize{\small}
\tablecaption{Comparison of Photometry to 2MASS Catalog.
\label{tab:2mass}
}

\tablehead{
\colhead{Date}       &
\colhead{Target}     &
\colhead{Passband}   &
\colhead{Magnitude\tablenotemark{(a)} }  &
\colhead{Number of}  &
\colhead{Median\tablenotemark{(b)} }     &
\colhead{$1\sigma$\tablenotemark{(c)}}   \\
\colhead{      }     &
\colhead{      }     &
\colhead{      }     &
\colhead{Range }     &
\colhead{Matches}    &
\colhead{Offset}     &
\colhead{Scatter}      
}

\startdata
14~Dec~2001 & IC~348 & $K$   & $ 11.00 - 13.75 $ & 280 & 0.075 & 0.051  \\
14~Dec~2001 & IC~348 & $H$   & $ 11.00 - 14.50 $ & 345 & 0.049 & 0.053  \\
14~Dec~2001 & IC~348 & $J$   & $ 11.00 - 15.50 $ & 343 & 0.050 & 0.067  \\
17~Dec~2001 & Off 1  & $K$   & $ 11.00 - 13.75 $ & 103 & 0.073 & 0.033  \\
08~Feb~2002 & IC~348 & $\Ks$ & $ 11.75 - 13.75 $ & 210 & 0.079 & 0.078  \\
08~Feb~2002 & IC~348 & $H$   & $ 12.00 - 14.50 $ & 267 & 0.086 & 0.066  \\
08~Feb~2002 & IC~348 & $J$   & $ 12.00 - 15.50 $ & 334 & 0.085 & 0.068  \\
11~Feb~2002 & Off 2  & $\Ks$ & $ 11.75 - 13.75 $ &  98 & 0.077 & 0.078 
\enddata

\tablenotetext{(a)}{For the magnitude range compared for the FLAMINGOS and 2MASS
   photometry of IC~348, the bright limit depended upon the source saturation in the
   FLAMINGOS images while the faint limit depended upon the increase in the RMS noise
   of the 2MASS data with magnitude.}
\tablenotetext{(b)}{ Median Offset = 2MASS - FLAMINGOS$_{absolute}$.}
\tablenotetext{(c)}{ Scatter measured after secondary offset removed.}

\end{deluxetable}


\clearpage

\begin{deluxetable}{lcccccccccccccc}
\tablewidth{0pt}
\tablecolumns{15}

\tabletypesize{\small}
\tablecaption{IC~348 Power-Law IMFs derived from Model KLFs
\label{tab:imf:fits} }

\tablehead{
\colhead{Cluster}         &
\colhead{$\mbox{N}_{\Gamma}$ } &
\colhead{$\mk$}           &
\colhead{$\mass/\solarmass$\tablenotemark{(b)}}    &
\colhead{$\chisq$}        &
\multicolumn{10}{c}{IMF Parameters\tablenotemark{(c)}} \\
\colhead{Region}          &
\colhead{\tablenotemark{(a)} }               &
\colhead{Limit}           &
\colhead{Limit}           &
\colhead{Prob.}           &
\colhead{$\Gamma_{1}$}    &
\colhead{$1\sigma$}       &
\colhead{$m_{1}$}         &
\colhead{$1\sigma$}       &
\colhead{$\Gamma_{2}$}    &
\colhead{$1\sigma$}       &
\colhead{$m_{2}$}         &
\colhead{$1\sigma$}       &
\colhead{$\Gamma_{3}$}    &
\colhead{$1\sigma$}       }

\startdata
cluster & 3  & $14.0$  & $0.040$ & $0.70$  & $-1.49$ & $0.30$ & $0.79$ & $0.25$ & $-0.19$ & $0.18$ & $0.089$ & $0.02$ & $+1.75$ & $0.53$ \\
cluster & 3  & $15.0$  & $0.025$ & $0.63$  & $-1.53$ & $0.28$ & $0.83$ & $0.23$ & $-0.23$ & $0.14$ & $0.086$ & $0.01$ & $+2.28$ & $0.22$ \\
cluster & 3  & $16.5$  & $\sim0.01$ & $<0.01$ & \nodata & \nodata & \nodata & \nodata &  \nodata & \nodata & \nodata & \nodata & \nodata & \nodata \\ \hline
core    & 2  & $14.0$  & $0.040$ & $0.93$  & $-1.43$ & $0.38$ & $0.56$ & $0.18$ & $+0.43$ & $0.25$ & \nodata & \nodata & \nodata & \nodata  \\
core    & 2  & $15.0$  & $0.025$ & $0.83$  & $-1.37$ & $0.35$ & $0.47$ & $0.13$ & $+0.59$ & $0.17$ & \nodata & \nodata & \nodata & \nodata  \\
core    & 2  & $16.5$  & $\sim0.01$ & $0.37$  & $-1.47$ & $0.38$ & $0.61$ & $0.19$ & $+0.32$ & $0.19$ & \nodata & \nodata & \nodata & \nodata  \\ \hline
halo    & 2  & $14.0$  & $0.040$ & $0.76$  & $-0.72$ & $0.14$ & $0.10$ & $0.02$ & $+1.98$ & $0.46$ & \nodata & \nodata & \nodata & \nodata  \\
halo    & 2  & $15.0$  & $0.025$ & $0.41$  & $-0.75$ & $0.16$ & $0.10$ & $0.01$ & $+2.25$ & $0.26$ & \nodata & \nodata & \nodata & \nodata  \\
halo    & 2  & $16.5$  & $\sim0.01$ & $<0.01$ & \nodata & \nodata & \nodata & \nodata &  \nodata & \nodata & \nodata & \nodata & \nodata & \nodata \\ \hline
halo    & 3  & $14.0$  & $0.040$ & $0.81$  & $-1.23$ & $0.39$ & $0.83$ & $0.28$ & $-0.55$ & $0.14$ & $0.093$ & $0.01$ & $+1.89$ & $0.45$ \\
halo    & 3  & $15.0$  & $0.025$ & $0.40$  & $-1.25$ & $0.39$ & $0.82$ & $0.29$ & $-0.53$ & $0.14$ & $0.092$ & $0.01$ & $+2.20$ & $0.27$ \\
halo    & 3  & $16.5$  & $\sim0.01$ & $<0.01$ & \nodata & \nodata & \nodata & \nodata &  \nodata & \nodata & \nodata & \nodata & \nodata & \nodata \\ \hline \hline
Trap    & 3  & \nodata & $0.025$ & $0.99$  & $-1.21$ & $0.18$ & $0.60$ & $0.16$ & $-0.15$ & $0.17$ & $0.120$ & $0.04$ & $+0.73$ & $0.20$ \\
\enddata

\tablenotetext{(a)}{Number of power-laws, $\Gamma_{i}$, used in the underlying IMF of the model KLFs.}
\tablenotetext{(b)}{Conversion of the ${\mk}_{,\,limit}$~to $\mass/\solarmass$ using the 2e6 Myr DM97 isochrone. For ${\mk}_{,\,limit}\,=\,16.5$,
    we converted to mass using the \citet{bur97} and \citet{bcah02} evolutionary models. }
\tablenotetext{(c)}{Units of IMF parametes: $\Gamma_{i}$ are slopes for an IMF defined as the number of stars per
    unit $\log{(\mass/\solarmass)}$; $m_{j}$ are mass breaks given in units of linear solar mass ($\mass/\solarmass$).}
\tablecomments{\chisq~probability calculated for best fit model KLF over the range from $\mk=8\;-\;{\mk}_{,\,limit}$.
    Average IMF parameters calculated within the 0.35 confidence contour.}
\end{deluxetable}


\clearpage


\figcaption[fig01.eps]{
   FLAMINGOS $JH\Ks$~infrared color composite image of IC~348. 
   For orientation, north is up while east is left and the field of view
   is  $20.5 \times 20.5 \arcmin$.  The bright blue star to the north by northwest
   is $\omicron$~Persi.  The cluster's interface with the Perseus Molecular Cloud
   is clearly outlined by a series of nebular features and enshrouded or heavily
   reddened sources along the southern edge of the image. The series of blue-green-red
   sources to the south by southwest of the cluster center is the asteroid 545 Messalina.
   Individual images were obtained 08 February 2002 with a resolution of 0.6\arcsec/pixel
   and a typical seeing of $1.6\arcsec$.
\label{fig:jhk_image}}

\figcaption[fig02a.eps, fig02b.eps]{
   Infrared color-magnitude diagrams for sources detected in the IC~348 FLAMINGOS wide-field images.  
   No radial or photometric criteria have been applied to the sources in these diagrams.
   The locations of the sources in these diagrams are compared to the location of the \citet{dm97}
   and \citet{bcah98} pre-main sequence isochrones for 2 and 5 Myr at 320 pc.
   Reddening vectors \citep{coh81} with length $\av\,=\,7$ are drawn for 1.4, 0.08 and 0.02
   \solarmass~PMS objects at the cluster's mean age. 
   A) $J-H/H$ color-magnitude diagram containing 1534 sources (82\% of total);
   B) $H-K/K$ color-magnitude diagram containing 1739 sources (93\% of total).
\label{fig:cmag_all}}

\figcaption[fig03.eps]{
   Infrared color-color diagram for sources in the IC~348 FLAMINGOS region with $\mk < 15$.
   This diagram has not been filtered by photometric error and includes 97\%~(563/580) of
   the sources brighter than this $K$ magnitude limit.
   This observed color distribution is compared to the intrinsic colors of field dwarfs
   (O-M9, \citet[][, where the Bessell \& Brett colors where shifted by zeropoint offsets
   to the 2MASS system using the relations derived by Carpenter 2001)]{bb88,kirk00}),
   classical T-Tauri stars with optically thick disks \citep{mch97}, and giants. $\sim12\%$
   of the sources display excess infrared emission in this diagram.  A reddening vector of
   length $\av\,=\,7$ illustrates the modest reddenings seen by the majority of the cluster.
\label{fig:jhk}}


\figcaption[fig04.eps]{
   Radial Profile of the IC~348 Cluster.
   Surface density (in stars per square degree) is measured in circular annuli, centering on
   the location of the LL95 sub-group IC348a.  Profiles are calculated using annuli of equal
   area (histogram with bins of decreasing width) and annuli with equal radial steps
   (heavy solid line with error bars).  The width of the first annuli was $R_{0} = 1.75\,\arcmin$.
   Profiles are compared to the unreddened background surface density
   (lightly shaded band; width = $2\times$ the $1\sigma$ deviation of background, see text)
   and the background surface density reddened by $\av\sim4$.  Also shown:
   the division of cluster into sub-regions (vertical dashed lines, see text);
   a 1/r profile fit to the entire cluster region (dot-dashed line; $\chisq\sim4.9\times10^{-4}$));
   and a King profile fit to the cluster core (dotted line; $r_{core}=0.25 \mbox{pc}$; $\chisq\sim1$).
   Note(s): bins with $R \ge 10.33\arcmin$ have been geometrically corrected to
   account for the survey boundaries and the conversion from angular to physical scale (upper x-axis)
   is calculated for a distance of 320 pc.
\label{fig:rad_prof}}

\figcaption[fig05a.eps, fig05b.eps]{
   Spatial distribution and surface density profile of IC~348 sources.
   A) Locations of sources in different  magnitude ranges are shown in an equatorial
   tangent projection of the wide-field FLAMINGOS IC~348 region. 
   Large filled stars correspond to sources with spectral types A5 and earlier
   including the B0III giant $\omicron$ Persi. Large filled circles
   correspond to sources $\mk < 10$, smaller filled circles, $10 <\,\mk\,< 15$, and
   small open circles, $\mk > 15$. 
   B)  The surface density of IC~348 sources is shown filtered by a Nyquist
   sampled box (width $=200\arcsec$, see upper left). Only sources $\mk\,< 15$
   are used and contours are given as multiples of the field star surface density
   derived by reddening the off-field by $\av\sim4.0$ or the same field star
   surface density shown as the darkly shaded band in Figure \ref{fig:rad_prof}
   (i.e., 2305 stars/sq. deg). In this figure, filled circles correspond to 
   sources displaying infrarec excess in Figure \ref{fig:jhk}.
   The $R < 5\arcmin$ and $R = 10.33\arcmin$ boundaries of the cluster sub-regions
   are shown as concentric circles. 
\label{fig:spatial}}

\figcaption[fig06a.eps,fig06b.eps]{
   Extinction Maps of the IC~348 FLAMINGOS Region.
   Individual source $\av$~estimates derived in Section \ref{sec:images:avmaps}
   are converted to an extinction map using a Nyquist sampled box filter
   (width $=200\,\arcsec$; see upper left; same as previous figure).
   Contours are in steps of $\av=1$ from $\av=1\mbox{ to }20$; we label the $\av=2,5,10$ contours.
   Left hand panel: $\av$~map derived from the bright, likely cluster members;
   Right hand panel: $\av$~map of the extinction seen by the fainter, likely background stars.
   See text for sample selection and explanation of reddening estimates.
\label{fig:av_map}}

\figcaption[fig07.eps]{
   Observed distribution of $H-K$ color and derived $\av$~distribution for sources in IC~348.
   Panels (A) and (B): the fractional distribution of $H-K$ color divided by sub-region.
   Panels (C) and (D): the probability distributions of $\av$~derived from de-reddening
   sources in the $(H-K)/(J-H)$ color-color diagram.
   In all panels, the distributions are divided into the results from magnitude limited
  ``bright'' and ''faint'' samples. See text for sample selection and explanation of de-reddening.
\label{fig:hk_av}}


\figcaption[fig08.eps]{
   Raw $K$ band luminosity functions of IC~348 separated by sub-region.
   The KLFs are scaled to stars per square degree and are compared to the
   observed field star KLF derived by combining the two off-cluster fields. 
   Both the cluster core and halo sub-regions appear to dominate the un-reddened
   off-field counts for $\mk < 15$. Error bars are $1\,\sigma$ counting statistics.
\label{fig:klfs}}

\figcaption[fig09.eps]{
   Field star correction to the raw IC~348 sub-region KLFs.
   Panels (A) and (B) compare the raw sub-region KLFs to the observed
   field star KLF convolved with the $\av$~distributions for background stars
   derived in Section \ref{sec:images:redden} and displayed in Figure \ref{fig:hk_av}.
   The reddening field star KLF(s) are scaled only the area of the cluster
   sub-region and subtracted to yield the differential KLFs displayed in panels (C) and (D). 
   The differential KLFs are significant for $\mk\,\le\,16.5$, below which we over-estimate
   the field star correction.
   Error bars are the square root of the sum of the observed counts 
   and the predicted field star counts for each bin.
\label{fig:diffs}}

\figcaption[fig10.eps]{
   Differential IC~348 KLFs by region.
   A) Comparison of the differential IC~348 KLFs for the two cluster sub-regions.
   B) Sum of sub-region differential KLFs into the composite IC~348 KLF.
   Panel B compares the sum of the sub-region differential KLFs from two different
   background corrections. The ``Red'' correction is the field star KLF reddened by 
   the EPDF for the background stars; the ``Blue'' correction is the field star
   KLF reddened by the EPDF for the bright cluster stars.
\label{fig:cluslf}}

\figcaption[fig11.eps]{
   The Star Formation History of IC~348.
   The star formation history of IC~348 from \citet{herbig98}
   is compared to that SFH used in our model luminosity function algorithm.
   The Herbig SFH is the merger of the ages derived for stars with and without detectable
   $\halpha$~emission using the de-reddened optical color-magnitude diagram.
   The SFH assumed for our models has a mean age, $\tau = 2.0$~Myr with
   constant star formation from 0.5 to 3.5 Myr ago.
\label{fig:sfh}}

\figcaption[fig12.eps]{
   Comparison of theoretical mass - apparent $K$ magnitude relations relevant for IC~348.
   Mass-$K$ magnitude relations from two different sets of theoretical evolutionary models
   and having different ages and distances are compared to illustrate the sensitivity
   of our method to assumptions about cluster distance and age.  Mass-$K$ magnitude relations
   from \citet{dm97} are shown at $\tau = $ 2,3 and 10 Myr for a distance of 320 pc and
   at $\tau = 3$ Myr for a distance of 260 pc.  Mass-$K$ magnitude relations from 
   the \citet{bcah98} tracks are also compared at 2 and 10 Myr at a distance of 320pc. 
   For all of these comparisons, the PMS tracks were converted to observables using a
   single tabulation of bolometric corrections.
\label{fig:kml}}

\figcaption[fig13.eps]{
   Fits of model KLFs to the composite IC~348 Cluster KLF and resulting cluster IMFs.
   The left hand panel compares the best fit model KLFs to the composite differential IC~348 KLF. 
   The right hand panel compares the IMFs of these model KLFs to the IMF derived
   for the Trapezium cluster by \citet{aam02}, although the Trapezium is only the
   $R < 0.3$ pc core of the larger Orion Nebula Cluster \citep{lah97}.
   Model KLFs and IMFs for IC~348 have corresponding symbols.
\label{fig:klf_imf_clus}}

\figcaption[fig14.eps]{
   Best fit Model KLFs and underlying IMFs for the IC~348 core/halo sub-regions.
   Left hand panels display model KLFs best fit to the differential IC~348 KLFs
   of each sub-region (upper: core; lower: halo).
   The best fit model KLFs are normalized to the observations over a range from $\mk=8-14$,
   corresponding to a mass range from $2.5\mbox{ to }0.04\solarmass$.
   Right hand panels display the underlying two and three power-law IMFs
   corresponding to the specific model KLFs (by symbol).
   Derived IMF parameters are listed in Table \ref{tab:imf:fits}.    
\label{fig:klf_imf_rad}}

\figcaption[f15a.eps, fig15b.eps]{
   Comparison of the IC~348 and Trapezium $K$ band Luminosity Functions.
   Panel (A) compares the $K$ band LFs of IC~348 and the Trapezium,
   shifted to absolute magnitudes. No reddening corrections have been included but
   the Trapezium KLF has been scaled to contain the same number of stars as IC~348.
   Panel (B) compares the IC~348 differential KLF to two model KLFs representing the
   Trapezium evolved to the age of IC~348. The models use the star forming history and reddening
   for IC~348 but substitute the Trapezium IMF derived in MLLA02 and use two different sets
   of PMS tracks \citep[][DM97 and Bur97]{dm97,bur97}. This illustrates the predicted location
   and size of the secondary KLF peak of the Trapezium were this cluster the age of IC~348.
\label{fig:trap_ic348}}

\figcaption[f16.eps]{
   Comparison of the Core and Halo Sub-Region IMFs for IC~348.
   Shown are Monte Carlo simulations of the derived IMFs for the IC~348 sub-regions
   and calculated for a sample of 150 stars.  Error bars, calculated as the $1\sigma$ variation
   in each IMF bin from 100 iterations, are shown every 0.3 dex in log mass for clarity.
   These sub-region IMFs are also compared to two published derivations of the IC~348a sub-cluster
   derived by \citet{ntc00} using NICMOS narrow band imaging and by \citet{luh2000} using a
   merger of optical and infrared spectra with infrared colors. Both sample somewhat
   smaller areas and contain fewer sources than the cluster ``core'' defined here and
   were scaled up in number for comparison.
   Error bars for the NTC00 study taken directly from that work. 
\label{fig:radimf}}


\newpage

\pagestyle{myheadings}
\pagenumbering{arabic}
\setcounter{page}{1}


\clearpage
\newpage
\markright{Figure 1}
\noindent 3 Color FLAMINGOS Image of IC~348.\\
Figure separately included as jpg (to make smaller).

\clearpage
\newpage
\setcounter{page}{2}
\markright{Figure 2a}
\begin{center}
\includegraphics[angle=00,totalheight=6.5in]{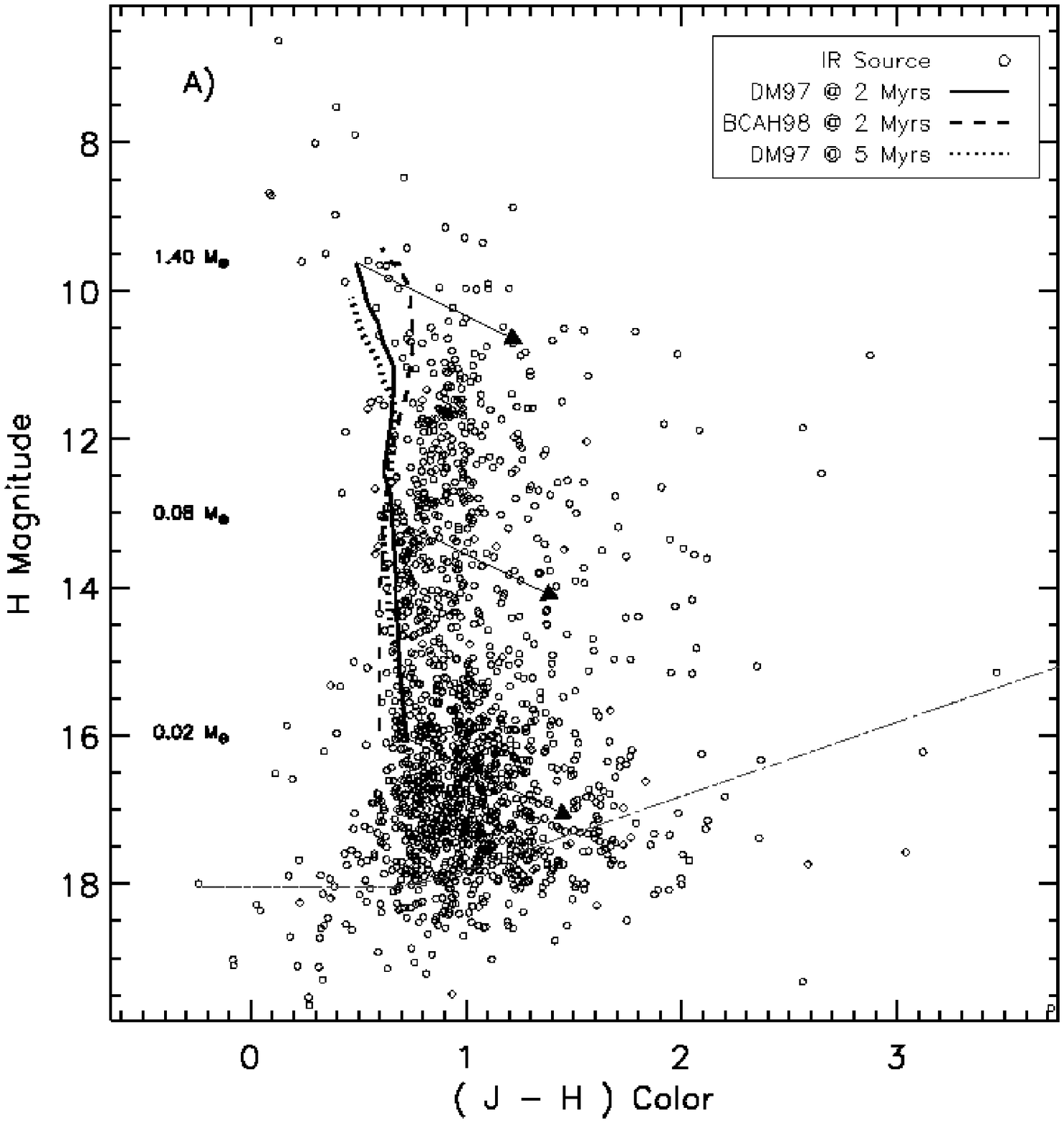}
\end{center}

\clearpage
\newpage
\setcounter{page}{2}
\markright{Figure 2b}
\begin{center}
\includegraphics[angle=00,totalheight=6.5in]{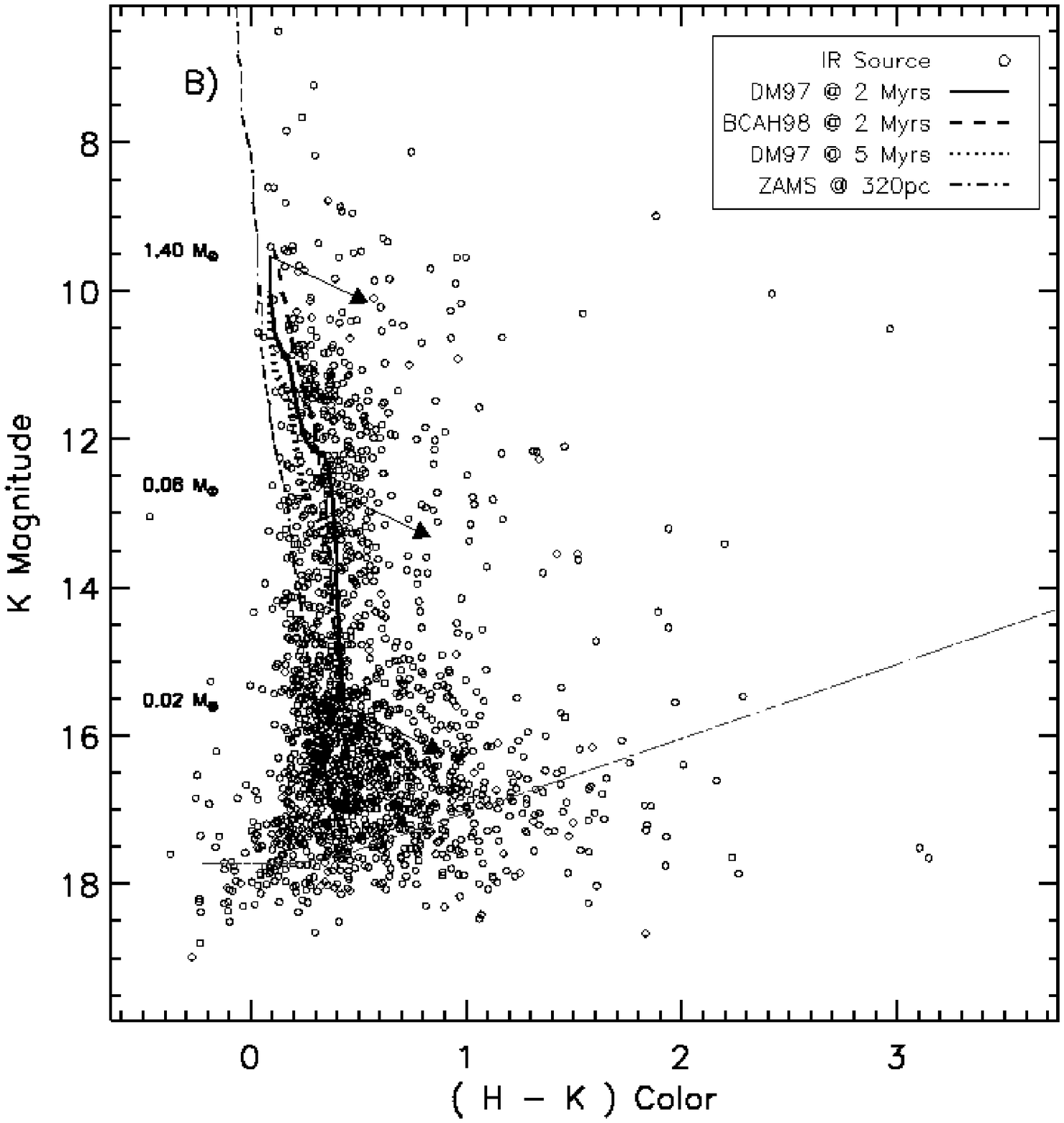}
\end{center}

\clearpage
\newpage
\markright{Figure 3}
\begin{center}
\includegraphics[angle=0,totalheight=6.5in]{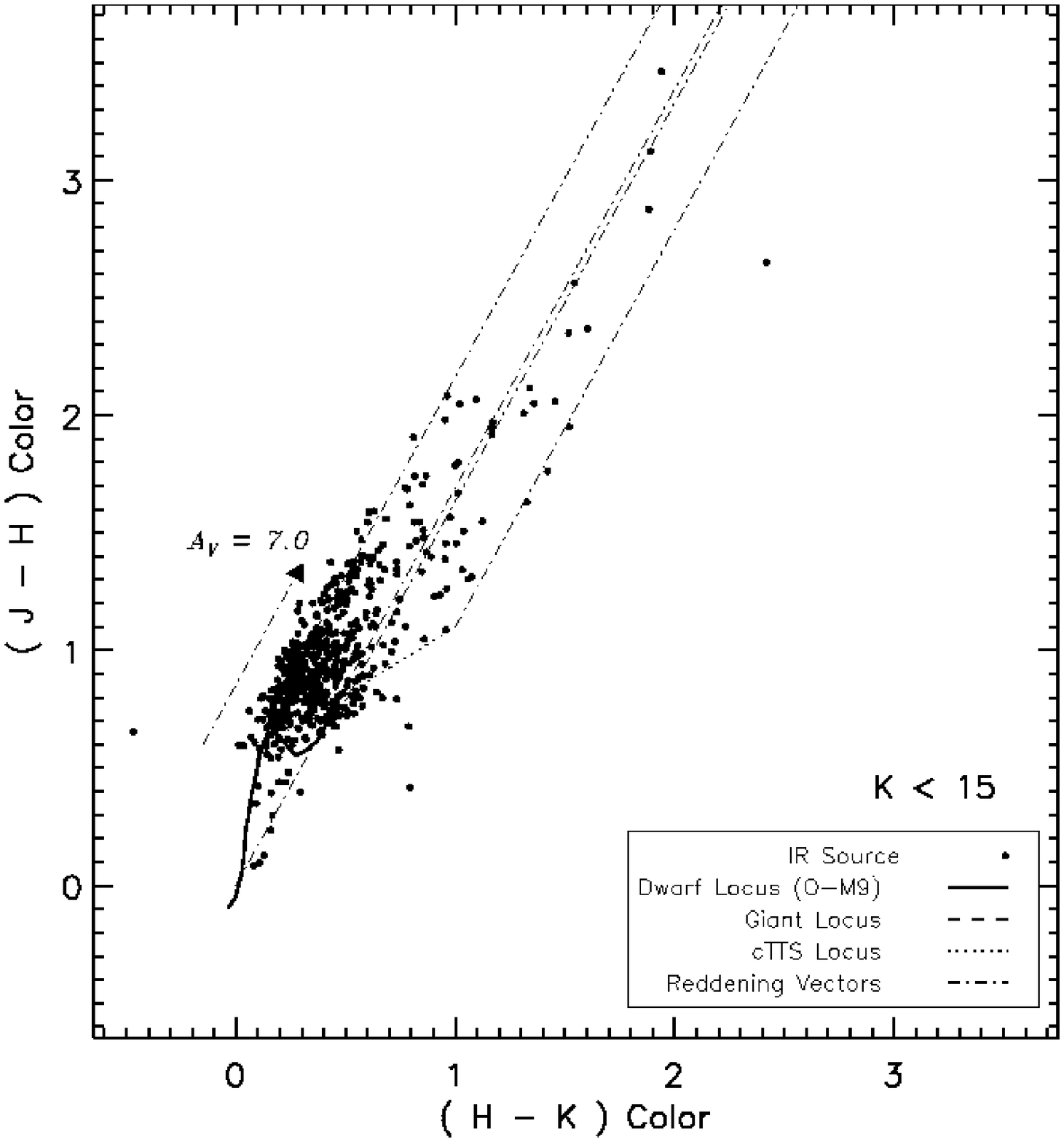}
\end{center}

\clearpage
\newpage
\markright{Figure 4}
\begin{center}
\includegraphics[angle=0,totalheight=6.5in]{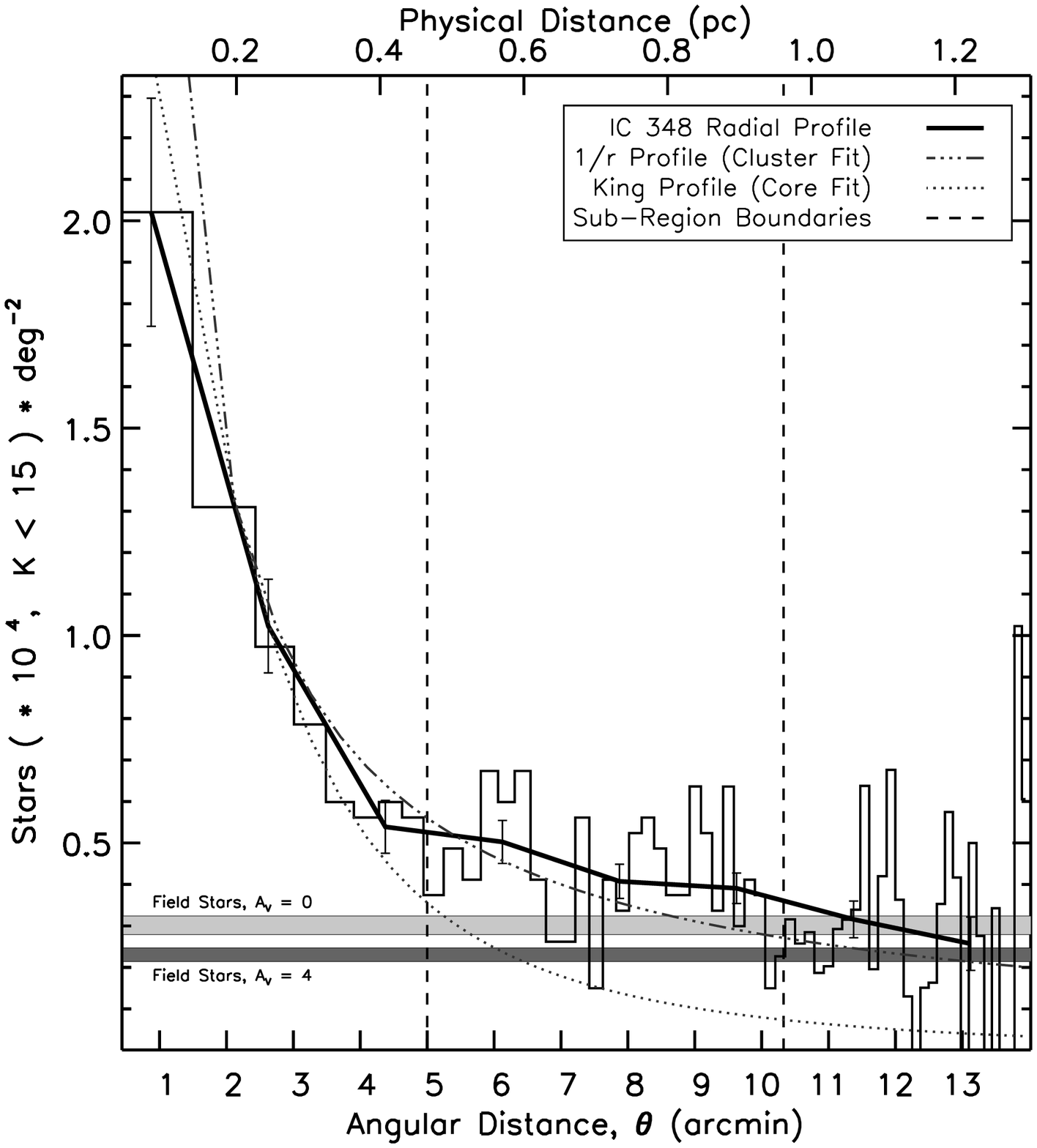}
\end{center}

\clearpage
\newpage
\setcounter{page}{5}
\markright{Figure 5a}
\begin{center}
\includegraphics[angle=00,totalheight=6.5in]{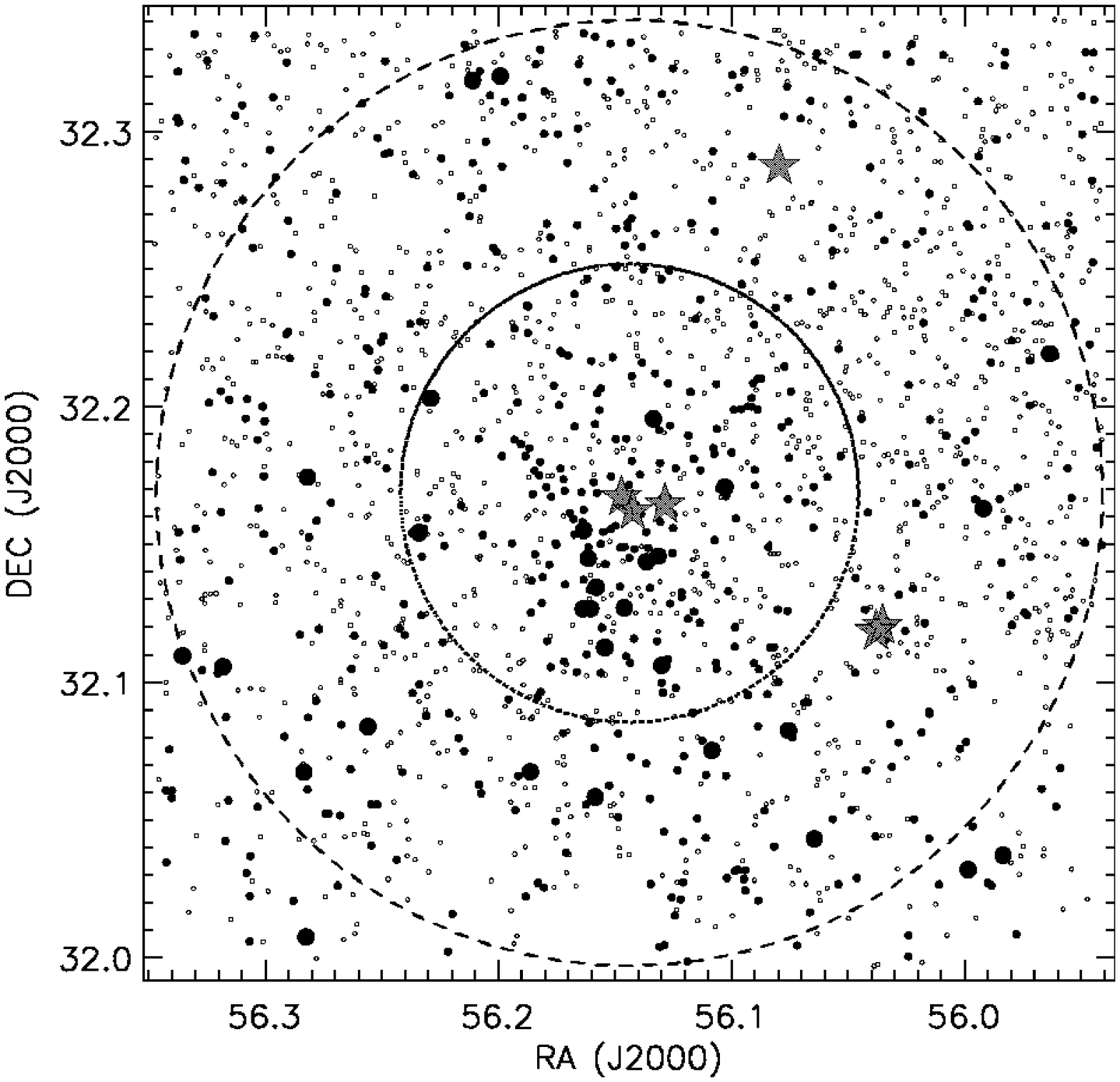}
\end{center}

\clearpage
\newpage
\setcounter{page}{5}
\markright{Figure 5b}
\begin{center}
\includegraphics[angle=00,totalheight=6.5in]{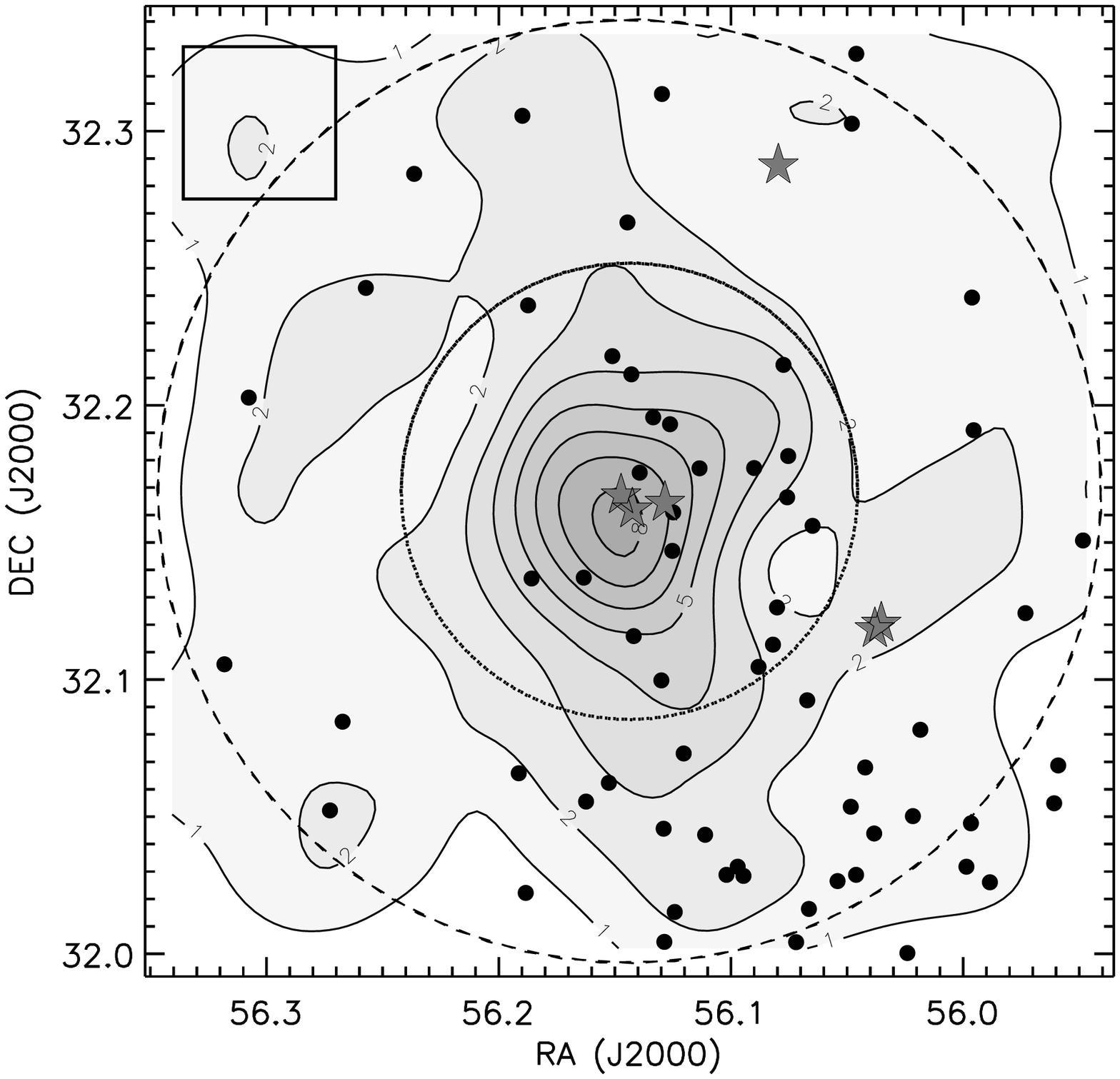}
\end{center}

\clearpage
\newpage
\setcounter{page}{6}
\markright{Figure 6a}
\begin{center}
\includegraphics[angle=00,totalheight=6.5in]{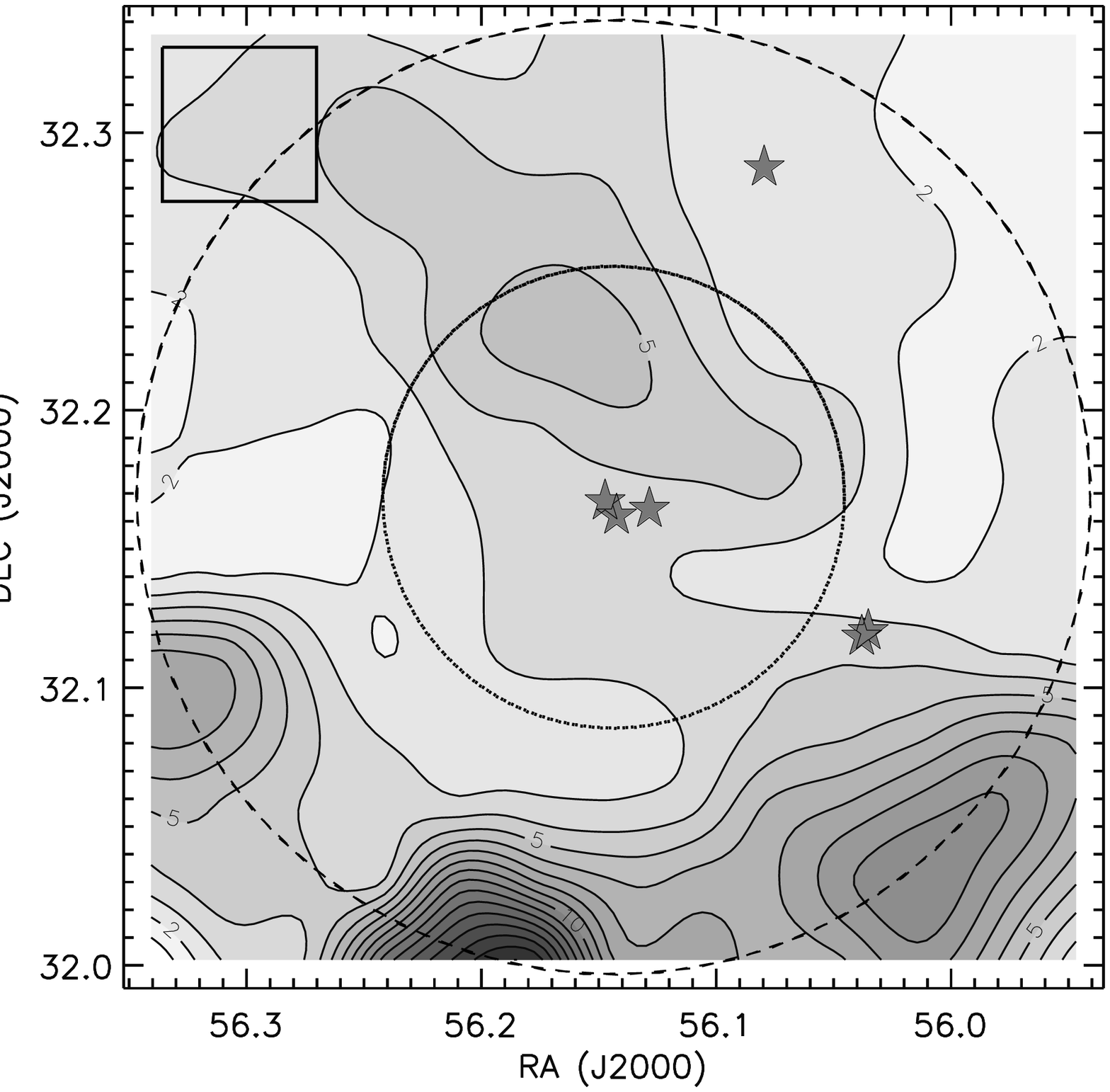}
\end{center}

\clearpage
\newpage
\setcounter{page}{6}
\markright{Figure 6b}
\begin{center}
\includegraphics[angle=00,totalheight=6.5in]{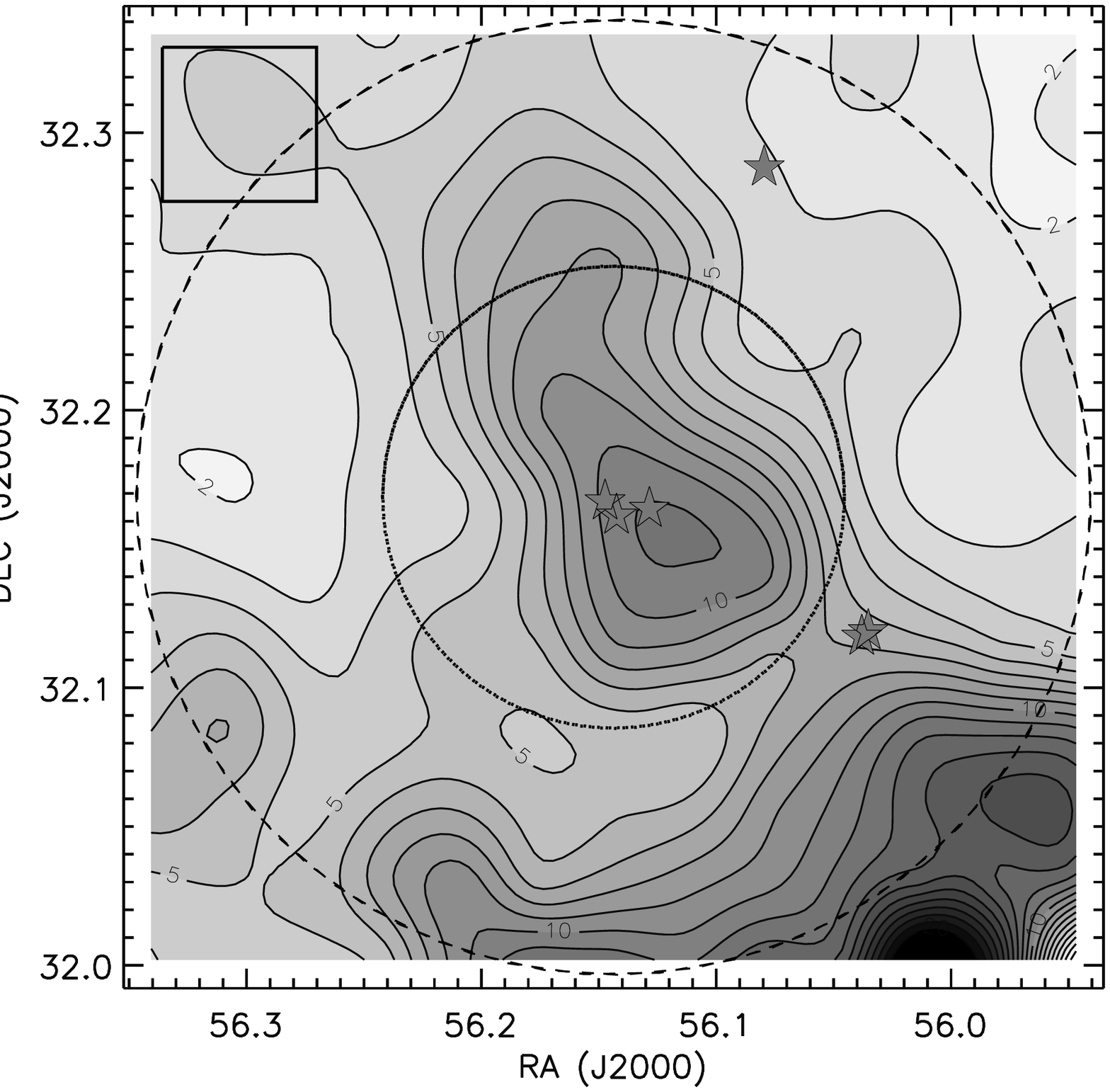}
\end{center}

\clearpage
\newpage
\markright{Figure 7}
\begin{center}
\includegraphics[angle=0,totalheight=6.5in]{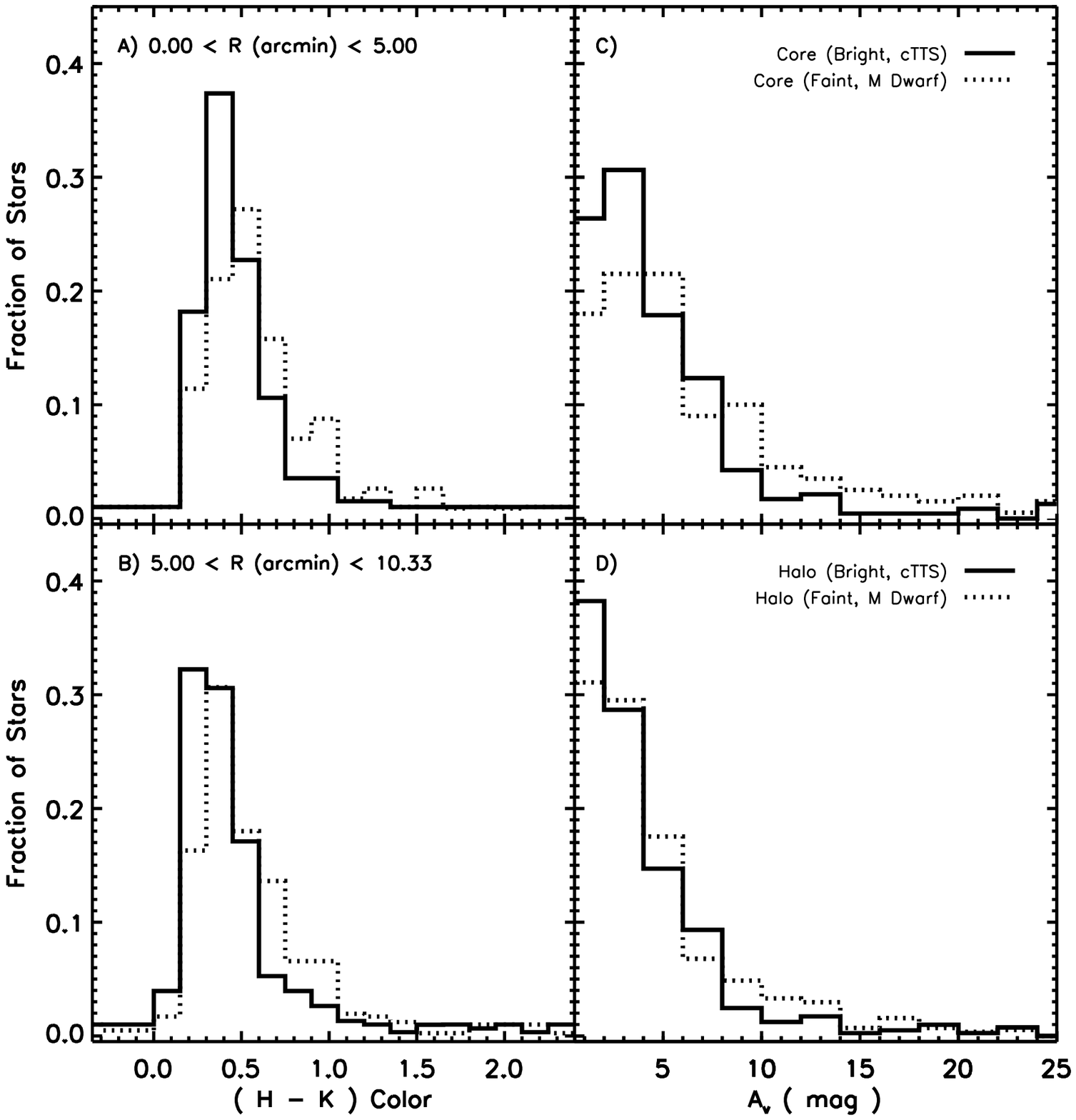}
\end{center}

\clearpage
\newpage
\markright{Figure 8}
\begin{center}
\includegraphics[angle=0,totalheight=6.5in]{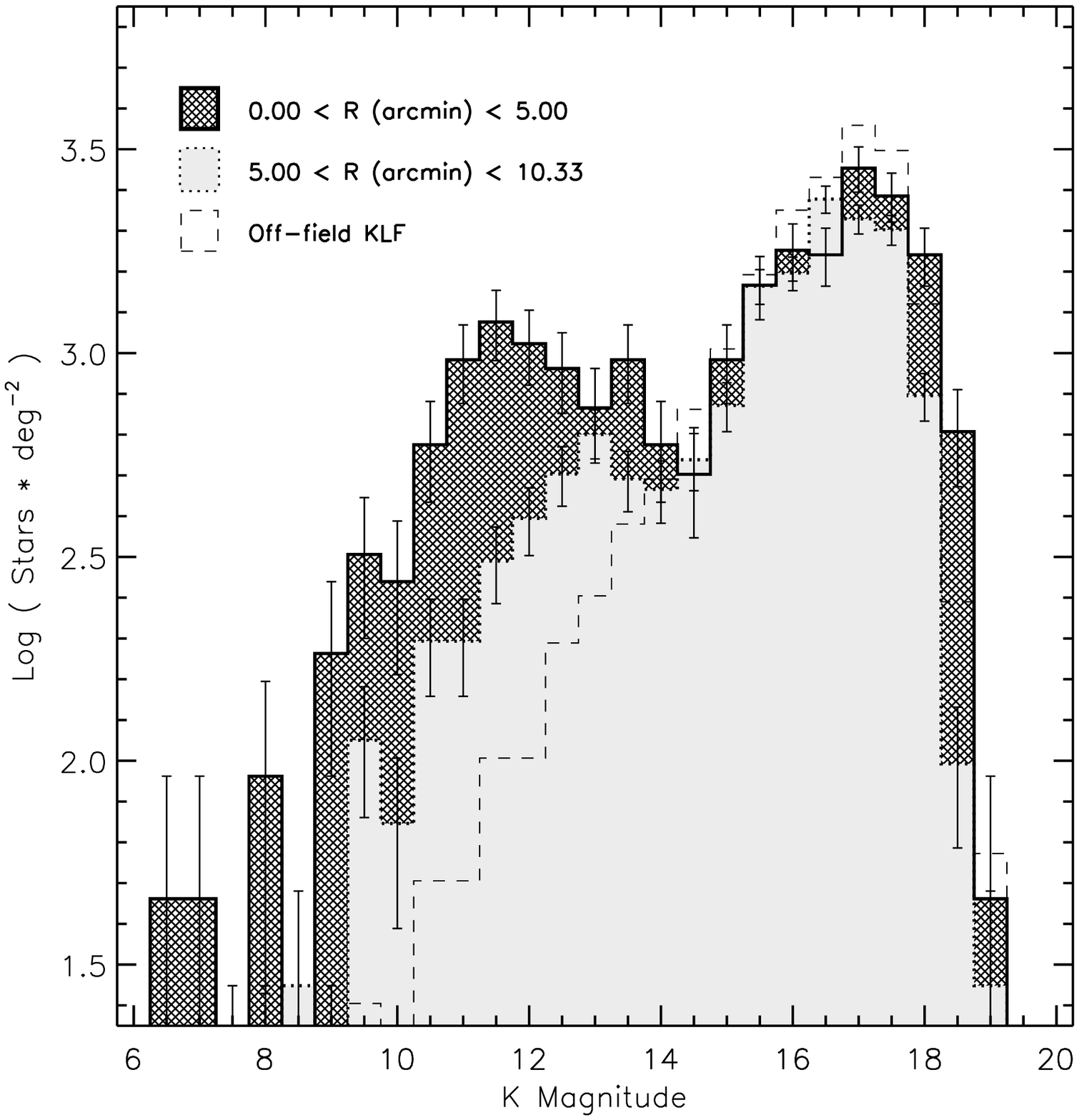}
\end{center}

\clearpage
\newpage
\markright{Figure 9}
\begin{center}
\includegraphics[angle=0,totalheight=6.5in]{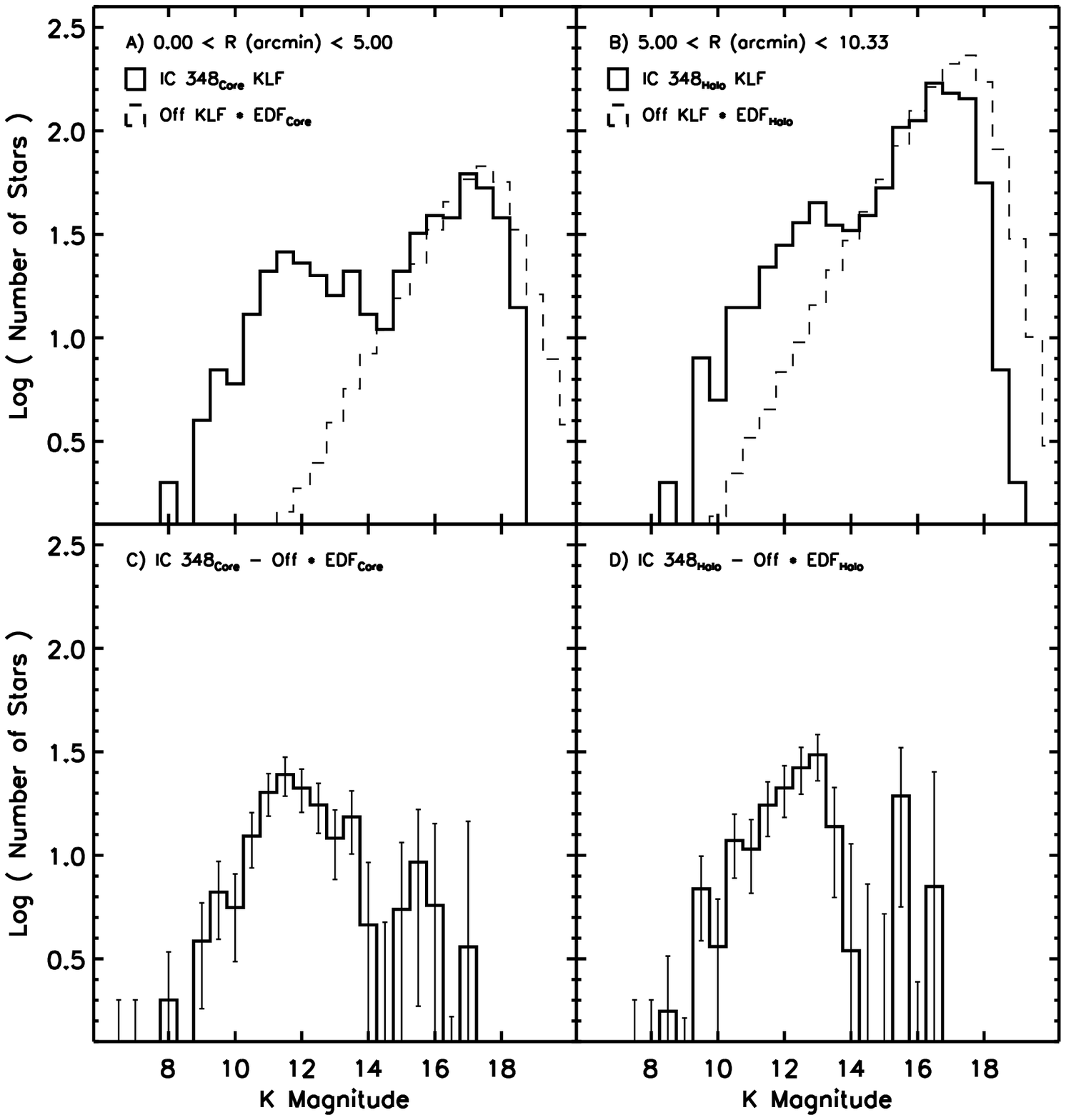}
\end{center}

\clearpage
\newpage
\markright{Figure 10}
\begin{center}
\includegraphics[angle=180,totalheight=6.0in]{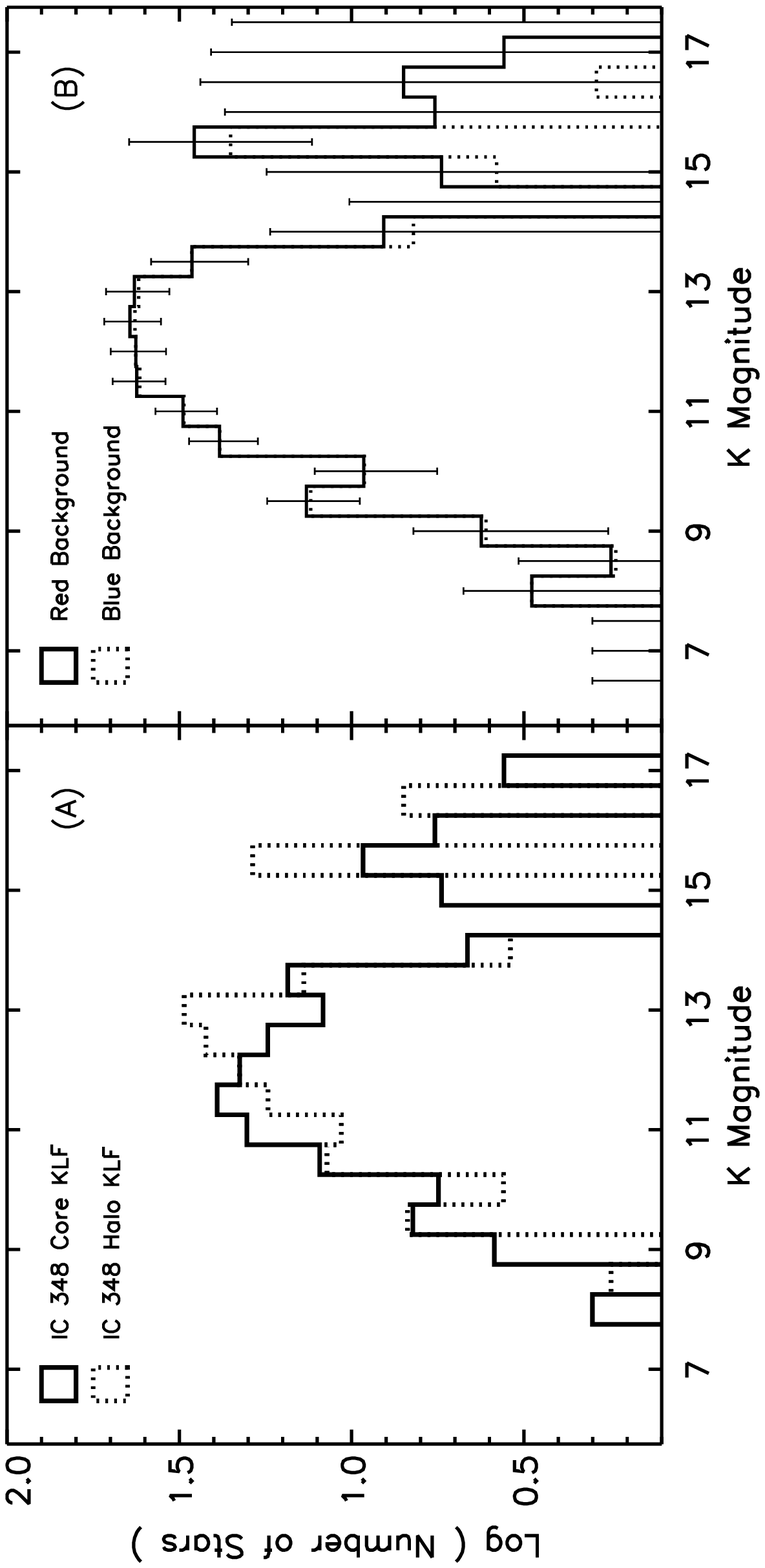}
\end{center}

\clearpage
\newpage
\markright{Figure 11}
\begin{center}
\includegraphics[angle=0,totalheight=6.5in]{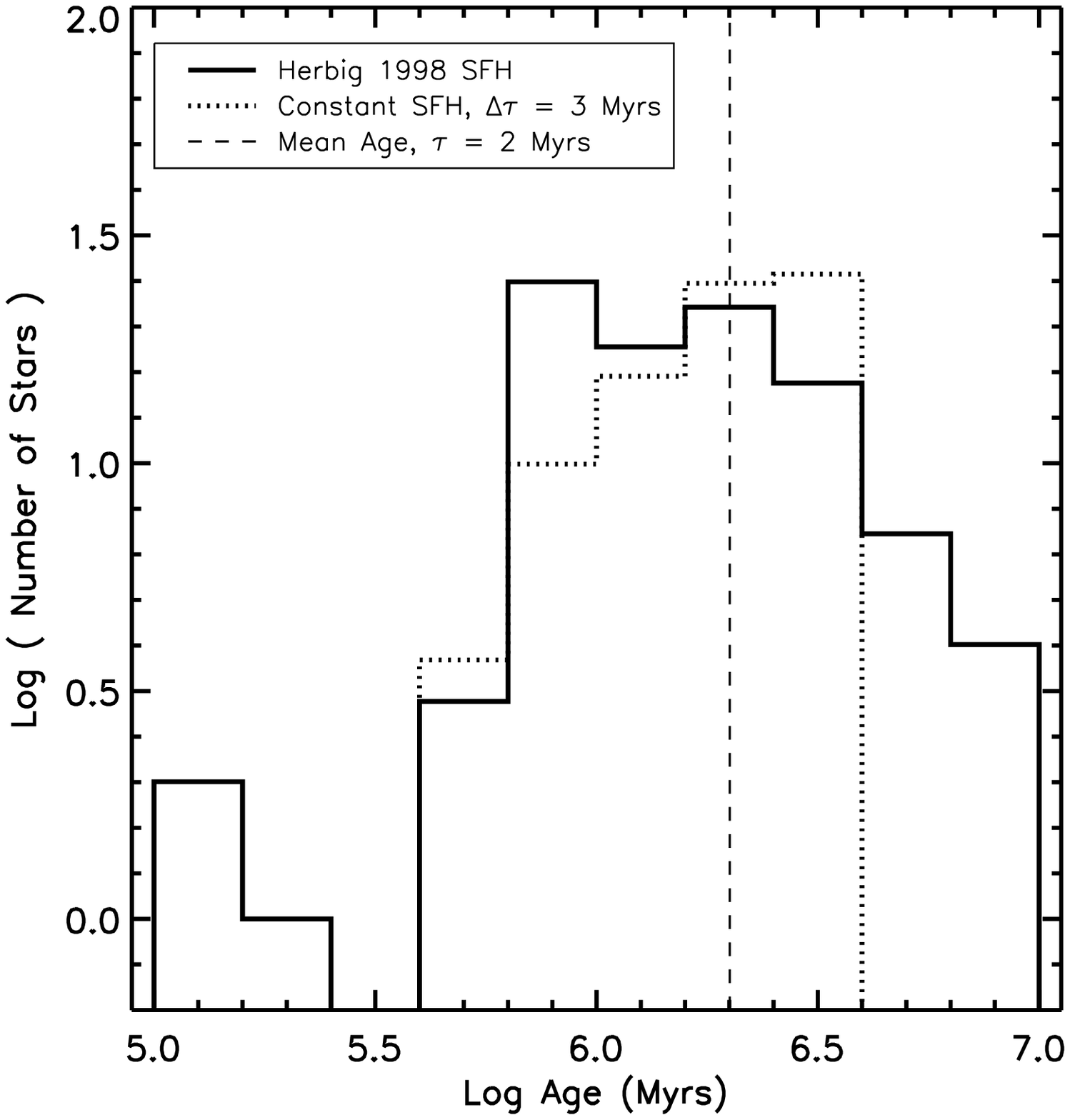}
\end{center}

\clearpage
\newpage
\markright{Figure 12}
\begin{center}
\includegraphics[angle=0,totalheight=6.5in]{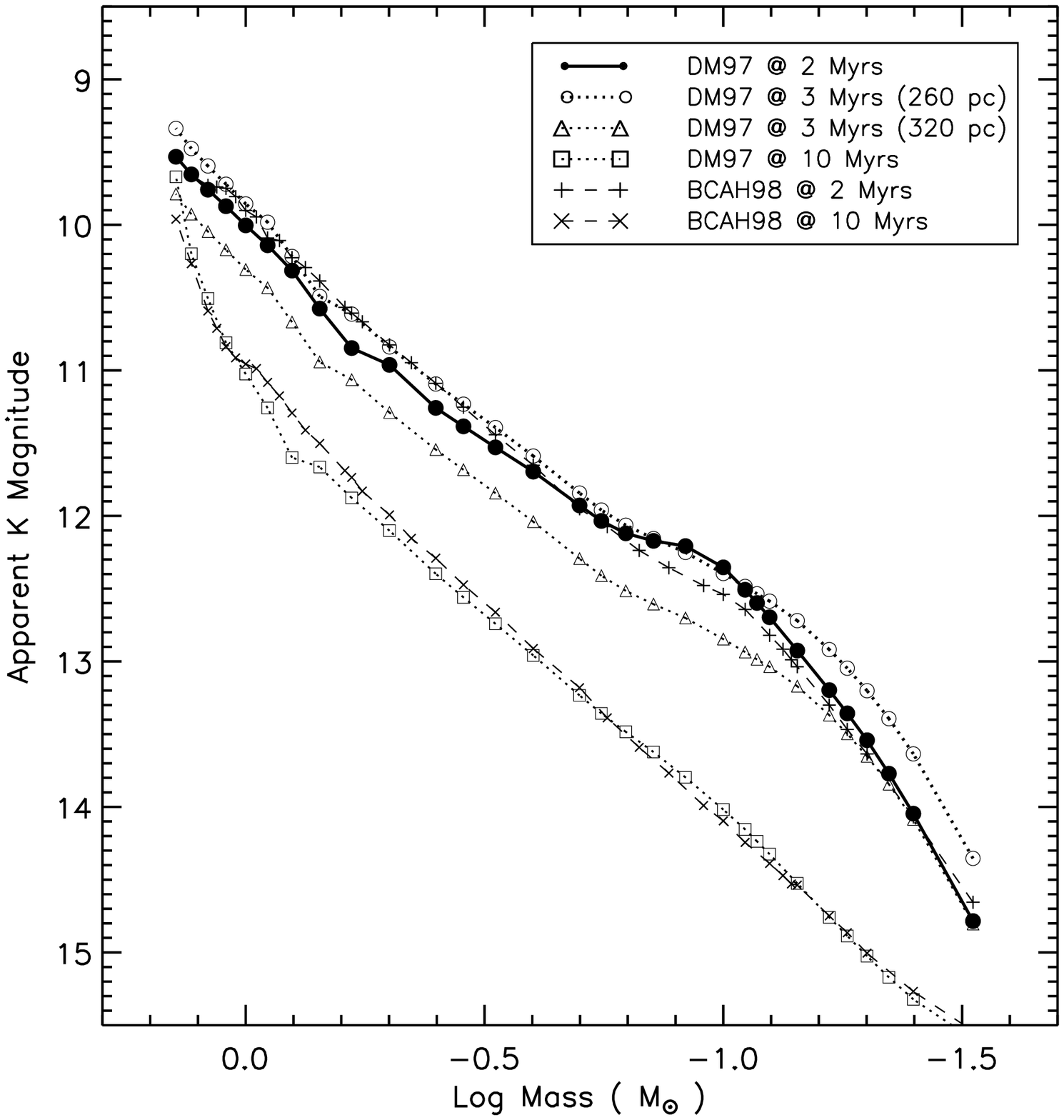}
\end{center}

\clearpage
\newpage
\markright{Figure 13}
\begin{center}
\includegraphics[angle=180,totalheight=6.0in]{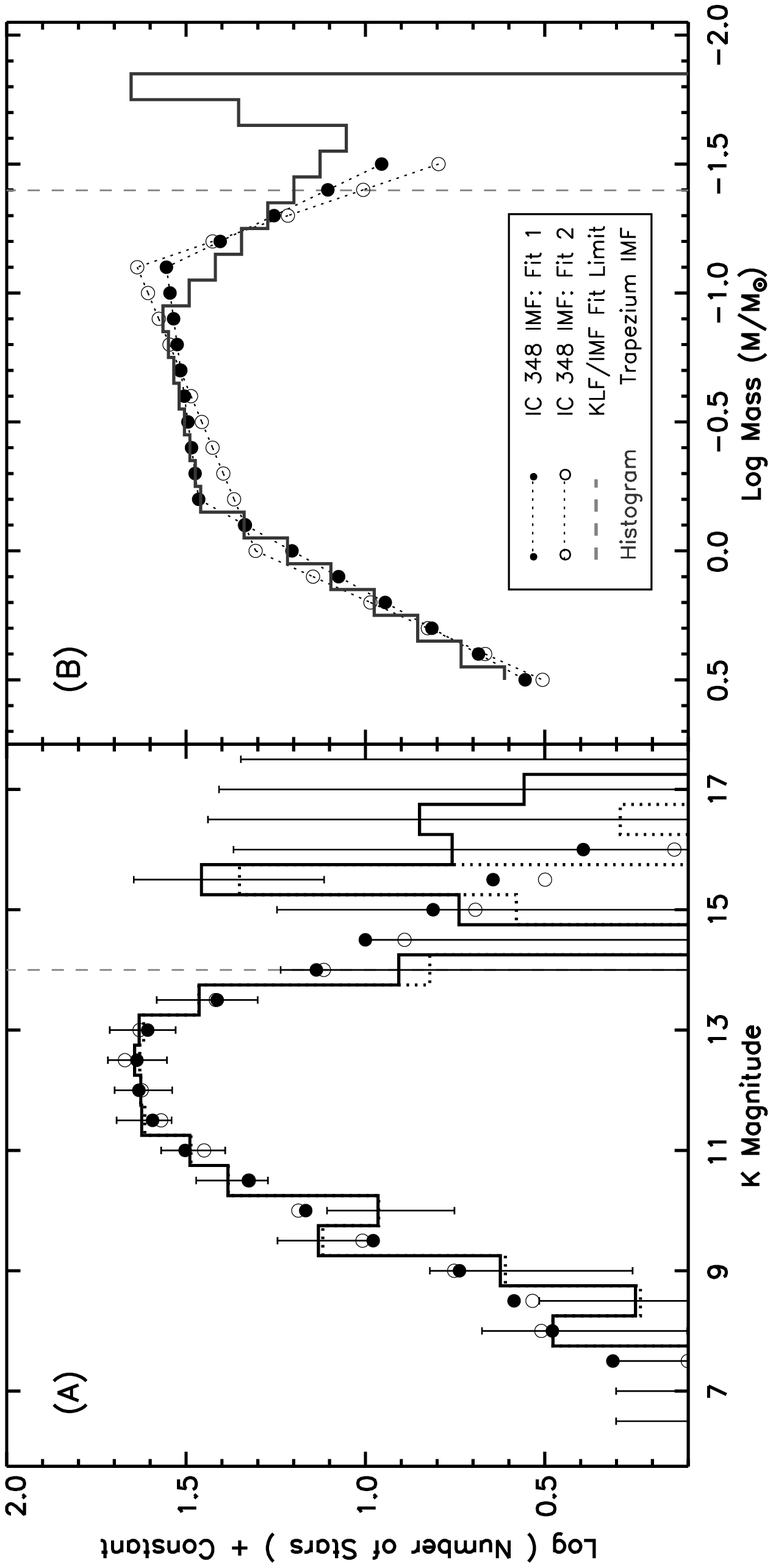}
\end{center}

\clearpage
\newpage
\markright{Figure 14}
\begin{center}
\includegraphics[angle=0,totalheight=6.5in]{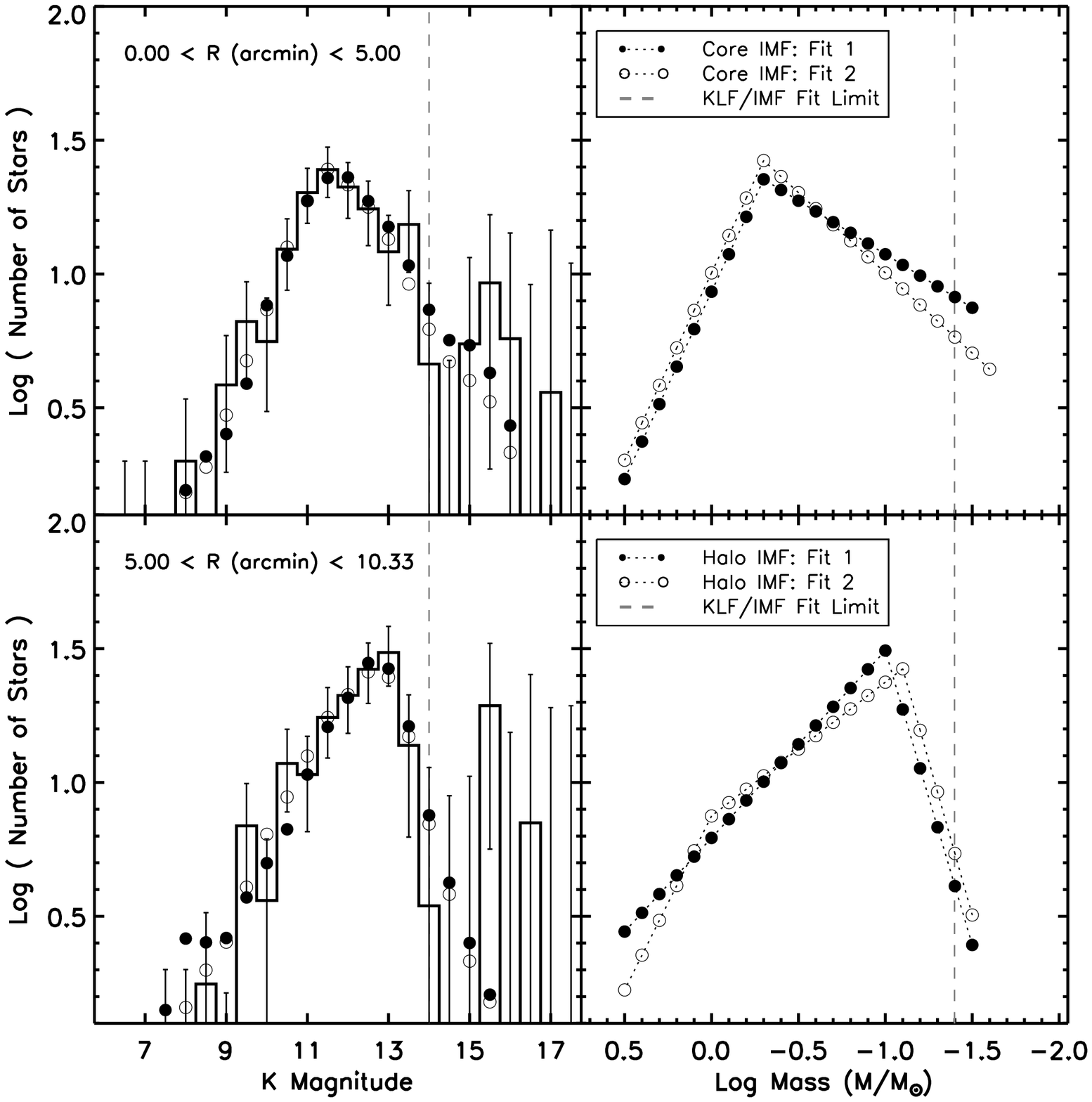}
\end{center}

\clearpage
\newpage
\markright{Figure 15a}
\setcounter{page}{15}
\begin{center}
\includegraphics[angle=0,totalheight=6.5in]{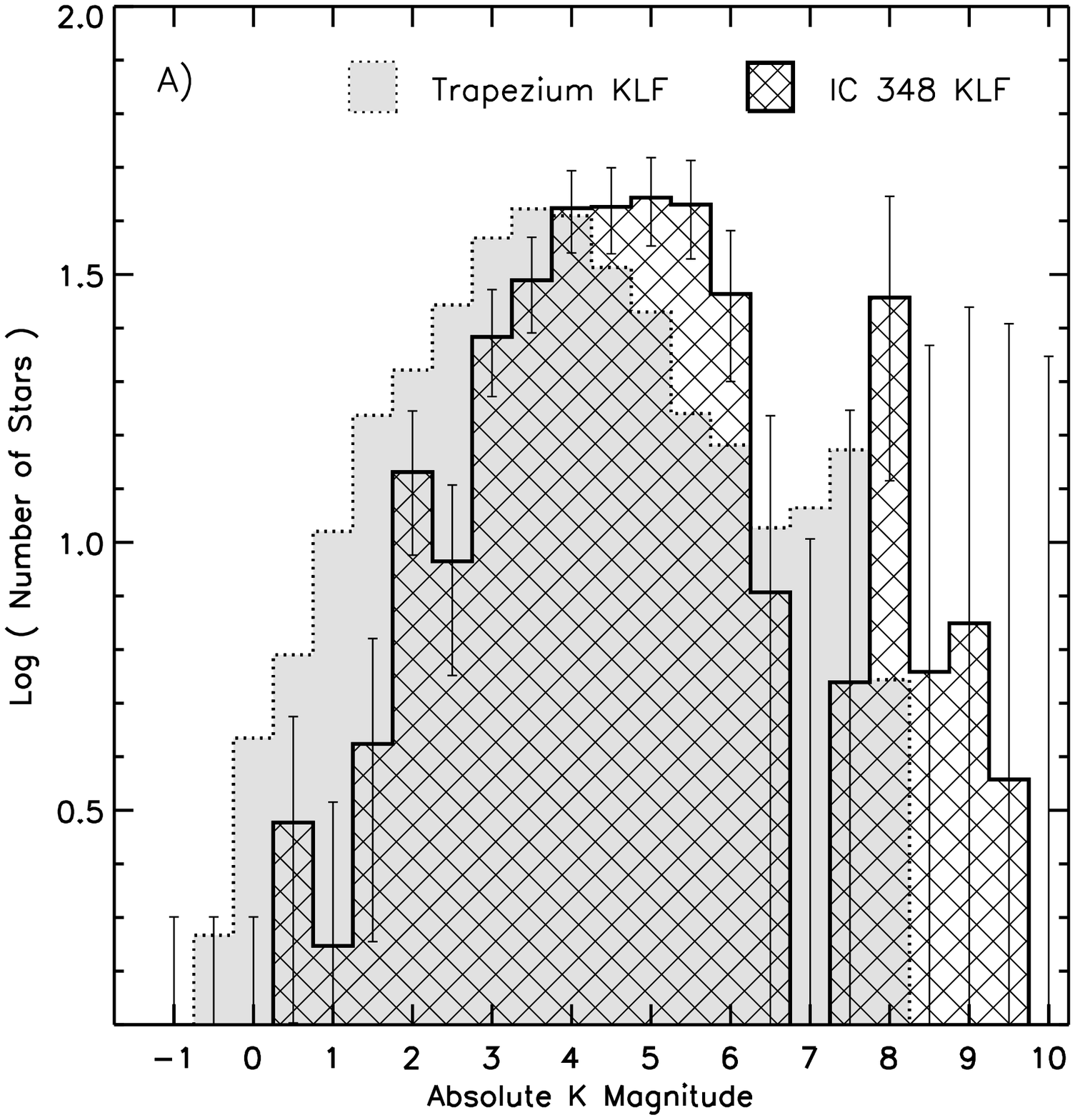}
\end{center}

\clearpage
\newpage
\setcounter{page}{15}
\markright{Figure 15b}
\begin{center}
\includegraphics[angle=0,totalheight=6.5in]{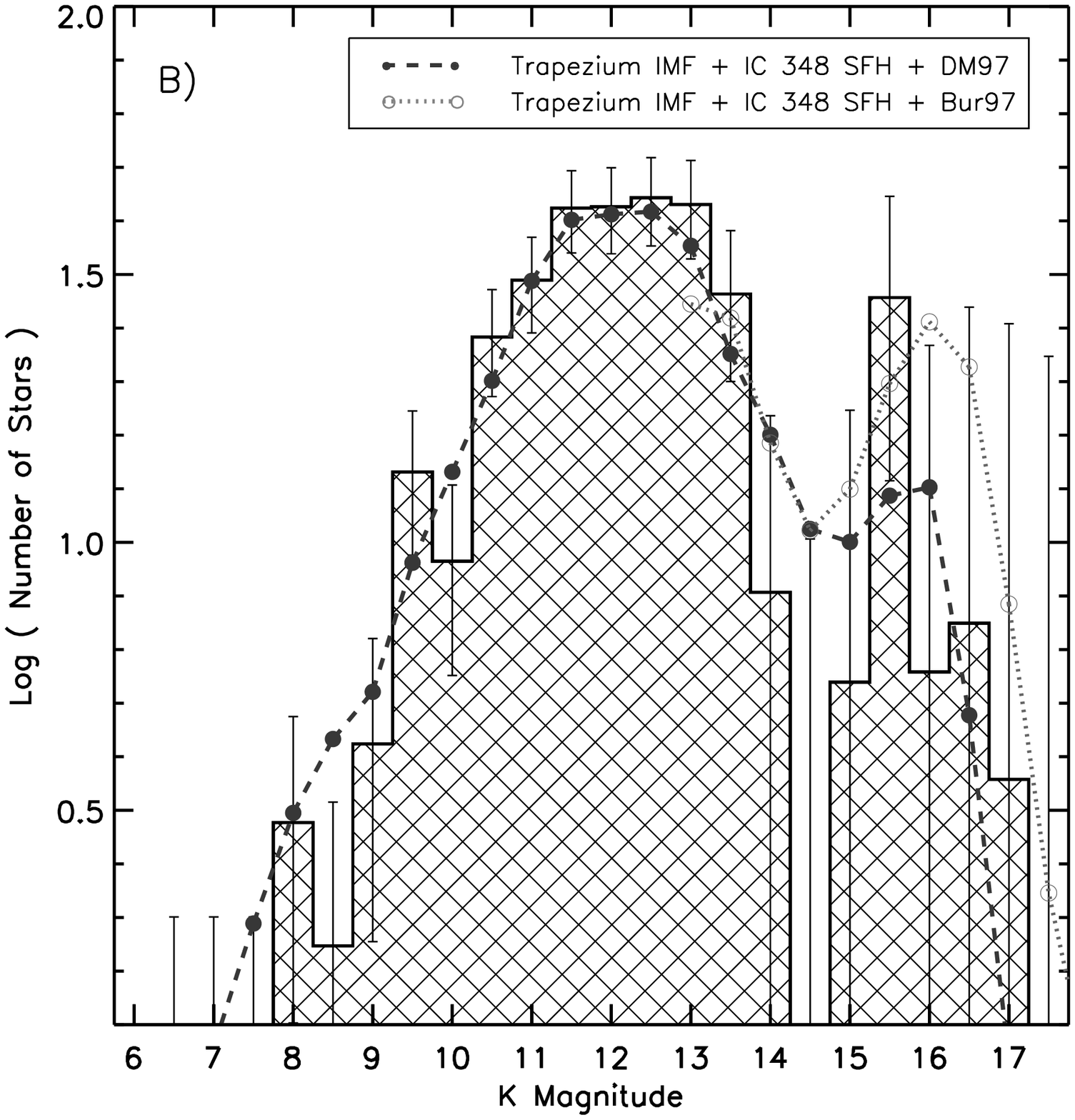}
\end{center}

\clearpage
\newpage
\setcounter{page}{16}
\markright{Figure 16}
\begin{center}
\includegraphics[angle=0,totalheight=5.75in]{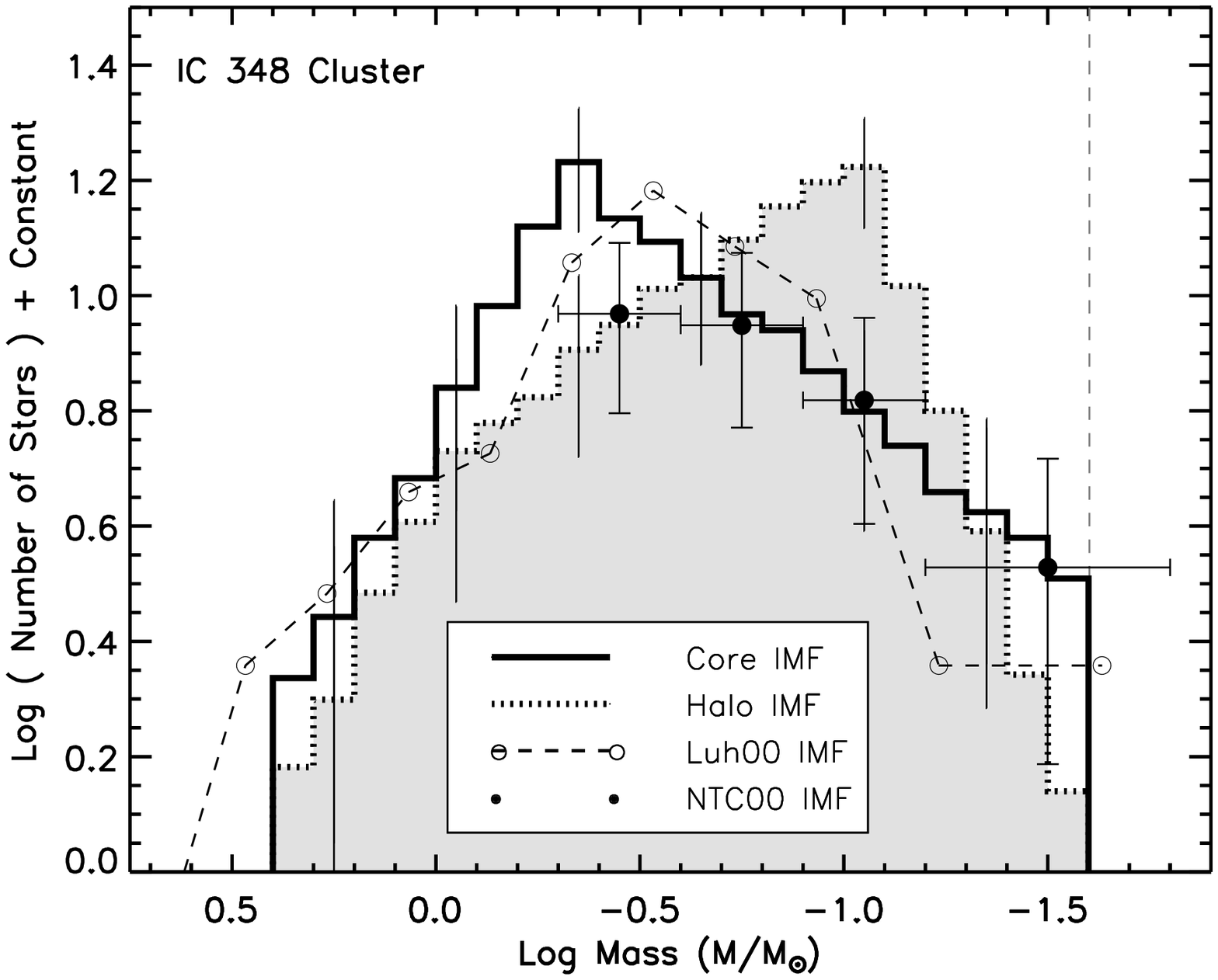}
\end{center}

\end{document}